\documentclass[a4paper,11pt]{article}

\usepackage[utf8]{inputenc}
\usepackage[english]{babel}
\usepackage{graphicx}
\usepackage{listings}
\usepackage{algorithm,algorithmic}
\usepackage{lmodern}
\usepackage[font=small]{caption}%
\usepackage{subcaption}
\usepackage{amsthm}
\usepackage{amsmath}
\usepackage{amssymb}
\usepackage{color}
\usepackage{relsize}
\usepackage{enumitem}
\usepackage{mathtools}
\usepackage{amsmath}
\usepackage{booktabs}
\usepackage{hyperref}
\usepackage{url}

\graphicspath{{./HP2C/}}

\begin{document}

\title{\vspace{-1.5cm}Study of fluid flow within the hearing organ}

\author{ Xavier Meyer$^{\text{\sf 1}}$, Elisabeth Delevoye$^{\text{\sf 2}}$ and Bastien Chopard$^{\text{\sf 1}}$\\
\small $^{\text{\sf 1}}$Department of Computer Science, University of Geneva, Switzerland. \\
\small $^{\text{\sf 2}}$CEA-Grenoble DRT/DSIS/SCSE, France. \\
%\url{bastien.chopard@unige.ch}\\
\normalsize
}

\date{}

\maketitle

%\part{Fluid simulation in the hearing organ}
\abstract{
Georg Von Békésy was awarded a nobel price in 1961 for his pioneering work on the cochlea function in the mammalian hearing organ~\cite{Bekesy1960}. He postulated that the placement of sensory cells in the cochlea corresponds to a specific frequency of sound. This theory, known as tonotopy, is the ground of our understanding on this complex organ. With the advance of technologies, this knowledge broaden continuously and seems to confirm Békésy initial observations. However, a mystery still lies in the center of this organ: how does its microscopic tissues exactly act together to decode the sounds that we perceive~?

One of these tissues, the Reissner membrane, forms a double cell layer elastic barrier separating two fundamental ducts of this organ. Yet, until recently~\cite{Reichenbach2012}, this membrane, was not considered in the modelling of the inner ear due to its smallness. Nowadays, objects of this size are at the reach of the medical imagining and measuring expertise~\cite{Shibata2009, Braun2012, Reichenbach2012}. Newly available observations coupled with the increasing availability of computational resources should enable modellers to consider the impact of these microscopic tissues in the inner ear mechanism.

In this report, we explore the potential fluid-structure interactions happening in the inner ear, more particularly on the Reissner membrane. This study aims at answering two separate questions :
\begin{itemize}
\itemsep0em 
\item Can nowadays computational fluid dynamics solvers simulate interaction with inner ear microscopic tissues~?
\item Has the Reissner membrane function on the auditory system been overlooked~?
\end{itemize} 

This report is organized as follow. Starting with a brief state of the art, the first section introduces the required notions to understand the experiments. The anatomy and the function of the cochlea are summarized and the main concepts of the computational fluid dynamics method used are defined. The next two sections presents the setting of the simulations and their results : the first focuses on the vestibular duct and the Reissner membrane while the second focuses on the cochlear duct and the organ of Corti. We conclude in the final section by discussing the numerical experiments results.}

\section{Introduction}

\subsection{State of the art}
Nowadays, more than 60'000 hearing impaired persons benefits from cochlear implants~\cite{Zeng2013}. These devices design is strongly tied to our understanding of the hearing organ. Its main component, the cochlea, processes acoustic signals by the mean of a complex mechanism. Since the early work of Georg Von Békésy~\cite{Bekesy1947, Bekesy1960}, research on the mammal cochlea has been the focus of many scientists~\cite{Ulfendahl1997, Robles2001}. Yet, this mechanism is not fully understood.

In order to fill this knowledge gap, a large amount of cochlear models have been developed through the years~\cite{Ni2014}. Given the complexity of the hearing organ, these models focus on a set of selected phenomenon : cochlear micro- or macro-mechanics with or without fluid coupling. Macro-mechanics models represent roughly the cochlea as a box with the vestibular and tympanic ducts separated by the vibrating basilar membrane~\cite{Elliot2012, Manoussaki2000, Zwislocki1974}. Micro-mechanics models focus on reproducing more accurately the organ of Corti~\cite{Bohnke1998, Geisler1995, Steele2009}. While all these models vary in the methods used and the size of phenomena simulated, they all start with the same assumption that the Reissner membrane does not play any major role in the mechanics of the cochlea~\cite{Ni2014}. However, this membrane, that separates two fundamental ducts of the cochlea, has been shown to propagate travelling waves in a comparable fashion as the widely considered basilar membrane~\cite{Reichenbach2012}.

To our knowledge, computational fluid dynamics (CFD) has only been used in one model~\cite{Givelberg2003}. This model presented an analysis of the macro-mechanics of the cochlea using immersed boundary conditions : a method simulating solid structure immersed in fluids. Our work is based on a similar simulation method but differs strongly in its aim. Our focus is the under-considered deformation of the Reissner membrane and its induced fluid displacement. We postulate that such fluid movements could impact the microscopic elements of Corti's organ and therefore play a key role in the hearing mechanism. In order to tackle this challenge of representing both the inner ear micro- and macro-mechanics in the same simulation setting, we use a highly-parallel CFD code, Palabos\footnote{http://www.palabos.org/}.

\subsection{Cochlea : the hearing organ}

\subsubsection{Anatomy}
\begin{figure}[ht]
  \center
  	\includegraphics[width=0.7\textwidth]{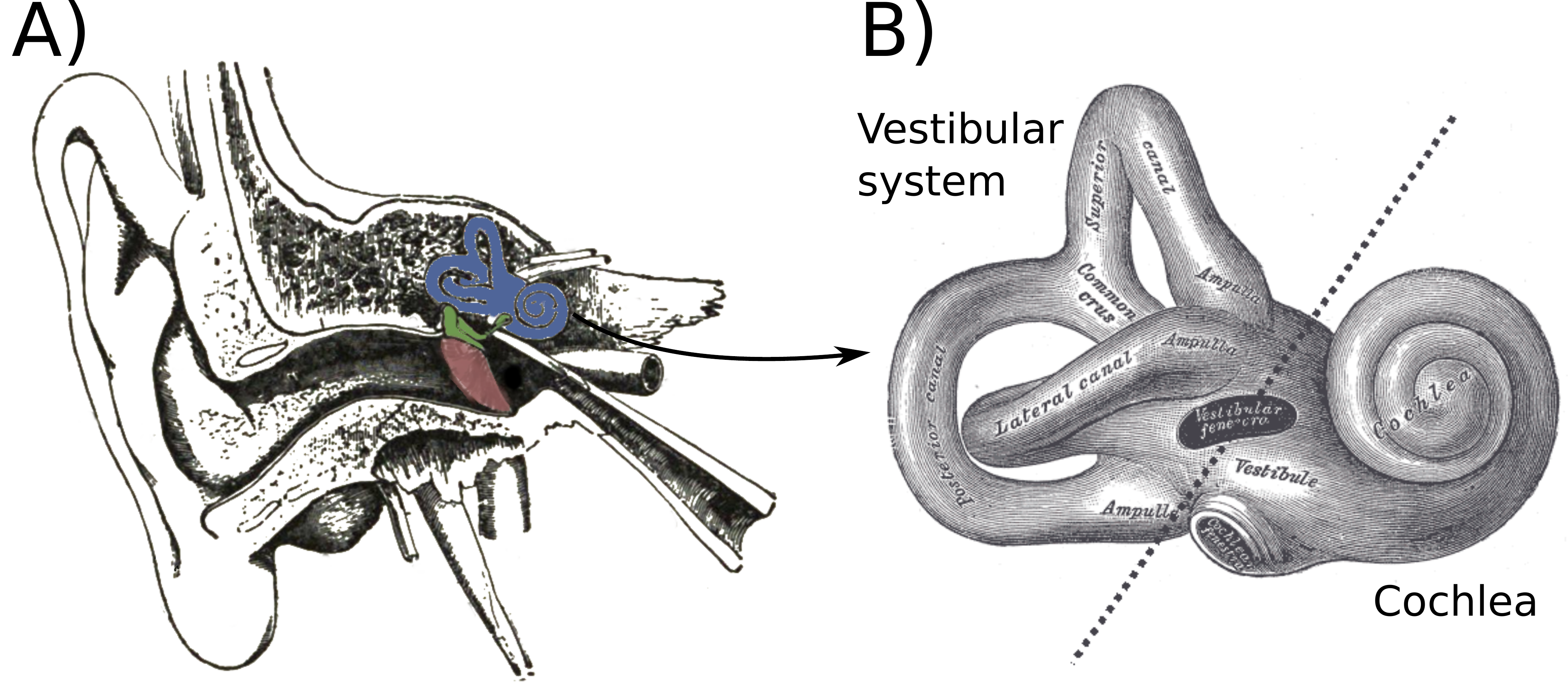}
  	 \captionof{figure}{Figure A) shows the chain of transmission of sound with the tympanic membrane (red), the ossicles (green) and the inner ear (blue). Figure B) shows the inner ear with its two main components : the vestibular system and the cochlea. Figures are adapted from~\cite{Gray1918}.}
  	 \label{fig::inner_ear}
\end{figure}

% Whole inner ear
The inner ear is composed of two major organ linked by the vestibule : the vestibular system and the cochlea (Fig.~\ref{fig::inner_ear}B). The main function of the first one is the sense of balance, while the second contains the primary auditory organ of the inner ear~\cite{Yost1989}.

\begin{figure}[ht]
  \center
  \includegraphics[width=0.7\textwidth]{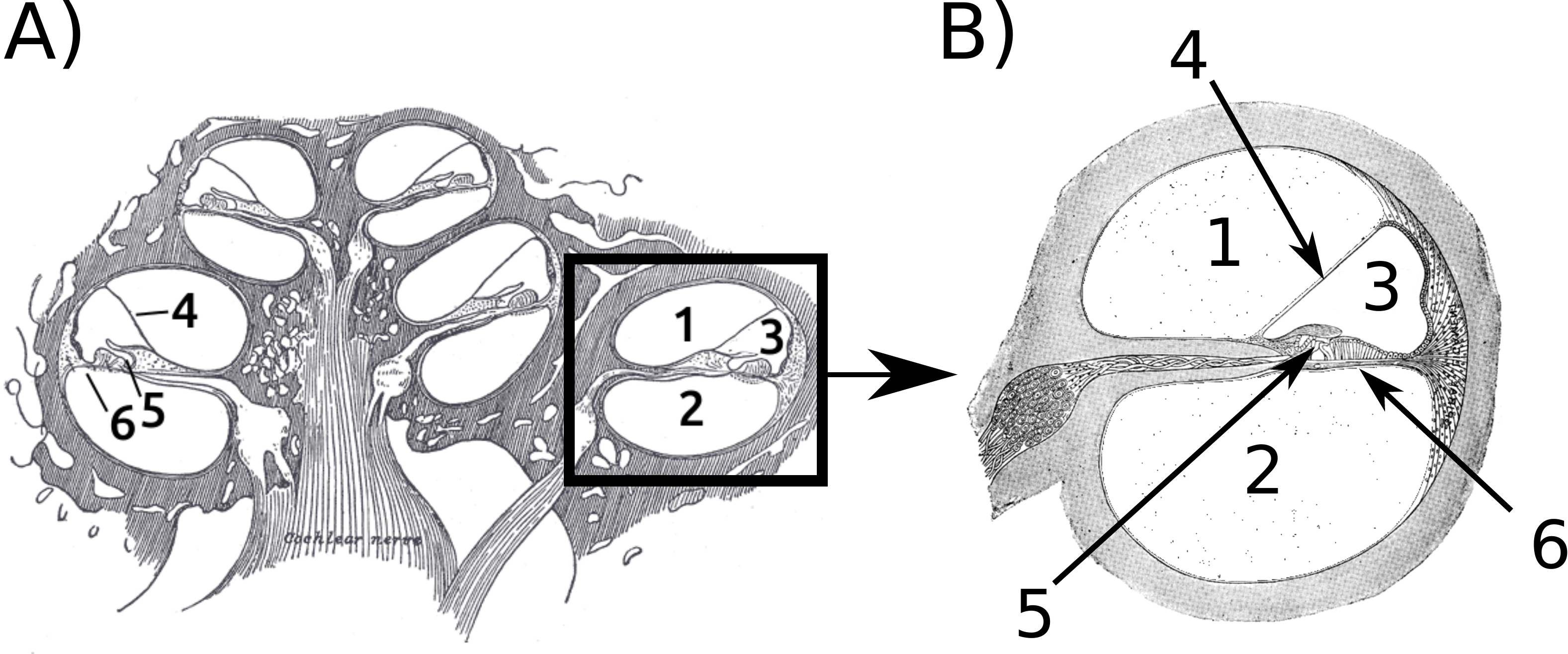}
  \captionof{figure}{Figure A) shows a cut of the cochlea and figure B) shows an enlargement of one cross-section of the ducts. The following elements are annotated : 1) Scala vestibuli; 2) Scala tympani; 3) Scala media; 4) Reissner membrane; 5) Corti's organ; 6) Basilar membrane. Figures are adapted from~\cite{Gray1918}.}
\label{fig::cochlea_cut}
\end{figure}

% Cochlea
The cochlea whose name comes from the ancient greek \textit{kohlias}, meaning spiral or snail shell, is structured as a spiral-shaped cavity. This cavity is considered to start from the base in the middle of the vestibule and continues until it reaches its apex, the helicotrema. In humans, this canal forms in average 2.5 turns over 35mm and is separated by a thin spiral shelf of bone, the osseous spiral lamina (Fig.~\ref{fig::cochlea_cut}A).

% Scala's and membrane
The cochlea can further be decomposed in three ducts separated by two membranes : the basilar and the Reissner membranes (Fig.~\ref{fig::cochlea_cut}B). One of these ducts, the vestibular duct, or scala vestibuli, starts from the vestibule and ends at the apex of the cochlea. The second duct, the tympanic duct, or scala tympani, starts from the tympanic cavity and connects with the vestibular duct at the cochlea apex through a small opening, the helicotrema. Both ducts contains a fluid, the perilymph that is sealed, on the opposite side of the helicotrema, by the oval window in the vestibule and the round window in the tympanic cavity.

In between both these ducts lies the cochlear duct or scala media. This duct contains a fluid, the endolymph and is separated, on one side, from the scala vestibuli by the delicate two-cells thick Reissner membrane. On the other side, it is separated from the scala tympani by the elastic basilar membrane. This membrane thickness reduces progressively from the base to the apex of the cochlea.

\begin{figure}[ht]
  \center
  \includegraphics[width=0.7\textwidth]{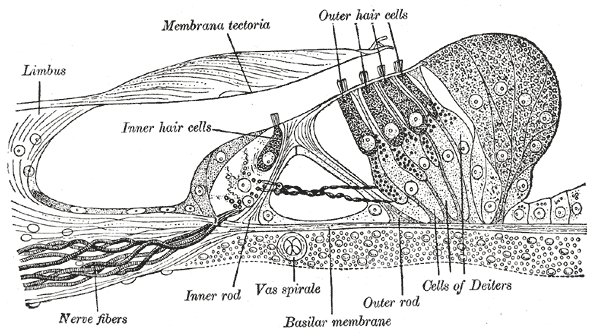}
  \captionof{figure}{Organ of Corti (adapted from~\cite{Gray1918}).}
  \label{fig:corti}
\end{figure}

Inside the scala media, on top of the basilar membrane rests the small yet important organ of Corti (Fig.~\ref{fig:corti}). This complex organ contains the fundamental element that translates the sound vibrations into nervous signals : the inner ear hair cells. These cells stand in a small canal, closed on one side by the basilar membrane and on the other side by the tectorial membrane.

% How sounds is translated ?
\subsubsection{The hearing mechanism}
Acoustic signals are conveyed in the inner ear by a vibratory pattern. Acoustic waves reach the external auditory canal, the tympanic membrane, and transmit vibrations through the ossicles, a complex of small bones (malleus, incus, stapes). These bones amplify the vibratory signal before applying it to the oval windows (Fig.~\ref{fig::inner_ear}A\footnote{The tympanic membrane or eardrum is represented in red and the ossicles in green.}). In turn, this membranous window transmits the vibrations to the vestibule and to the inner ear fluid. These vibrations in the fluids are considered to create a travelling wave on the basilar membrane. The wave evolve from the base toward the apex for a length inversely proportional to its frequency. Therefore the placement, along the cochlea, of inner ear hair cells defines the sound frequencies that activate them.

This theory, more known as tonotopy, was proposed by Georg Von Békésy~\cite{Bekesy1947, Bekesy1960} and is based on his observations of the basilar membrane vibratory response when excited by sounds. This explanation of the hearing mechanism grossly simplify the reality by hiding complex micro-mechanical phenomenon occurring in the organ of Corti. Moreover, it doesn't account of the role of the Reissner membrane. More complex hypothesis are still explored in order to explain the impact of the Reissner membrane~\cite{Reichenbach2012} or the role of the cochlea active mechanism that produce otoacoustic emissions~\cite{Kemp2002}.

% What kind of tool we use
\subsection{Computational fluid simulations}
\subsubsection{Palabos, an open source highly parallel solver}
%% 1) Palabos / LBM

% We want thus to simulate fluid into cochlea, or simplification of cochlea.
% Depending on geometry and settings can be rather power intensive
% Lattice Boltzmann -> Palabos

In order to simulate the complex fluid-solid phenomenons occurring in the inner ear, we used the lattice Boltzmann method (LBM). This method is a modern approach in Computational Fluid Dynamics. It is often used to solve the incompressible, time-dependent Navier-Stokes equations numerically. Its strength lies in the ability to easily represent complex physical phenomena, ranging from multiphase flows to chemical interactions between the fluid and the surroundings. The method finds its origin in a molecular description of a fluid and can directly incorporate physical terms stemming from a knowledge of the interaction between molecules.

Unlike traditional CFD methods, LBM models the fluid consisting of fictive particles, and such particles perform consecutive propagation and collision processes over a discrete lattice mesh. Due to its particulate nature and local dynamics, LBM has several advantages over other conventional CFD methods, especially in dealing with complex boundaries, incorporating microscopic interactions, and parallelization of the algorithm\footnote{http://en.wikipedia.org/wiki/Lattice\_Boltzmann\_methods}. For that matter, we chose ot use the open-source software Palabos\footnote{http://www.palabos.org/}. This software is a massively-parallel Lattice Boltzmann implementation that have been used in numerous applications such as the simulation of blood cells deposition in aneurysms~\cite{Zimny2013} or permeability change of porous medium~\cite{Huber2014}.

In order to simulate membranes, Palabos maintainers developed an elastic shell model. In this model, the fluid domain is bounded by time-independent boundaries (rigid walls) and time-dependent boundaries (membranes). During the simulation, the fluid-structure interaction is taken in account at these boundary such as to apply internal forces on the elastic wall. The time dynamics of the wall is then solved and the fluid domain is recomputed to adapt to these changes. This model implement all these steps as to be second-order accurate and offers a good scaling for parallel execution.

%% 2) String model 
\subsubsection{Palabos elastic shell model}

\begin{figure}[ht]
\centering
\includegraphics[width=0.8\textwidth]{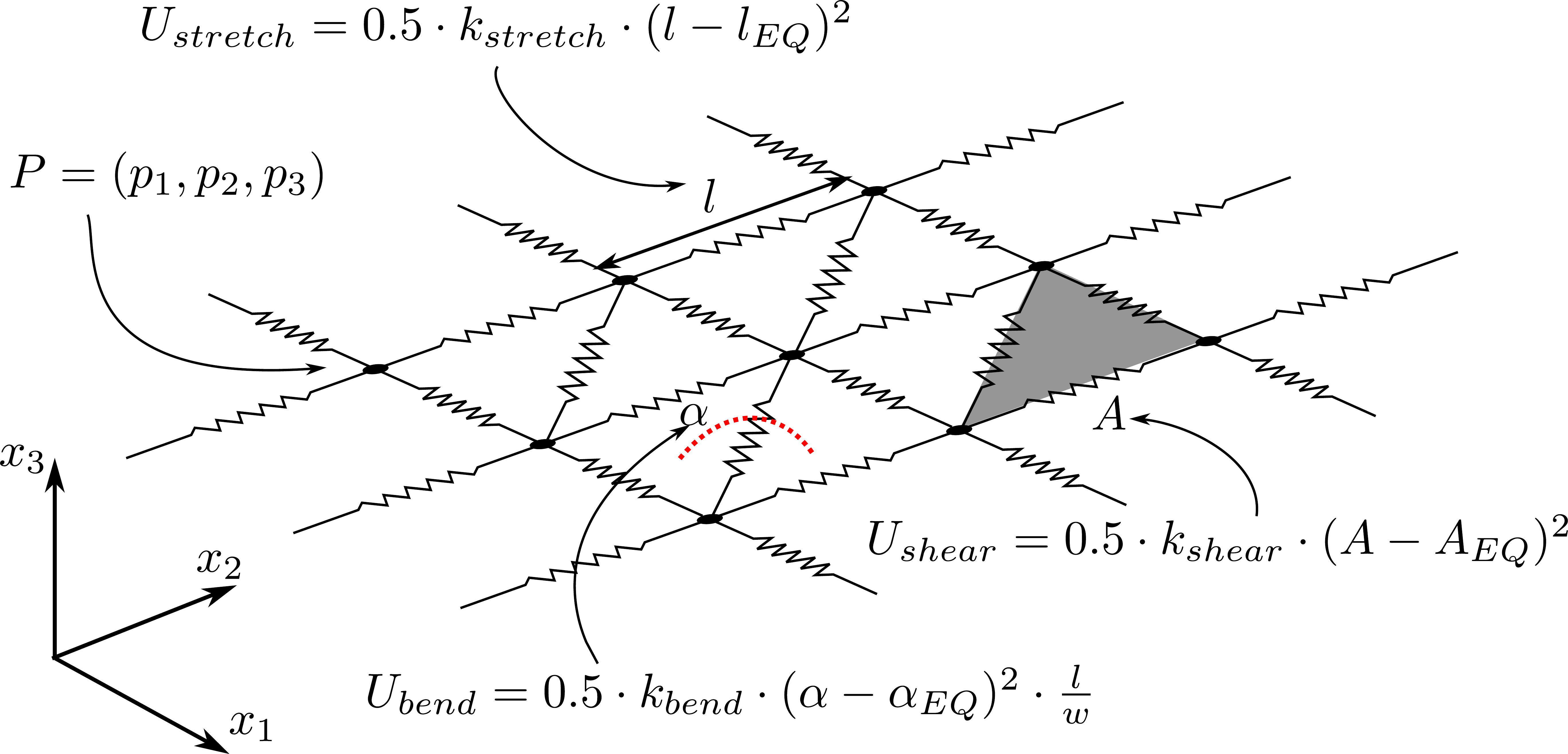}
\caption{Representation of a time-dependent boundary and the three forces acting on it : stretching, shearing and bending.}
  \label{fig:membrane}
\end{figure}

Time dependent boundaries, or membrane, are represented by triangular surfaces in a three dimensional space (Fig.~\ref{fig:membrane}). In order to simulate membrane movement, the surface vertices or membrane particles $P$ are displaced according to the stress generated by the fluid. This mechanism is guided by the simplified following iterative algorithm:
\begin{enumerate}
\item Fluid particles inside the fluid domain are detected and updated using the lattice-Boltzmann algorithm.
\item The fluid stresses on wall are computed and extrapolated to the membrane particles.
\item The elastic stress at each of these particles is then computed.
\item Membrane particles advance according to Newton's law.
\item The off-lattice boundary condition is reconstructed according to the new membrane position.
\end{enumerate}

The elasticity of the membrane is characterized by a complex model formed of stretching, bending and shearing forces (Fig.~\ref{fig:membrane}). The first force is modelled by considering the edges between vertex as spring with linear attractive forces. The second is insured by preserving the on-membrane angles. The latter is defined as the preservation of the triangular surfaces area.

At each iteration, these elastic forces are computed and used to represent the acceleration of each membrane particles. The elastic potential of a particle $U$ is represented by multiple potential in order to represent as realistically as possible the three dimensions. 

The stretch potential exists between a particle $P$ and each of its neighbours. This potential is given by
\[
U_{stretch} = 0.5 \cdot k_{stretch} \cdot (l-l_{EQ})^2
\]
where $l$ stand for the distance between the particles at this iteration and $l_{EQ}$ their distance at equilibrium.

The bending potential represent the angular elasticity of the membrane. For a given membrane particle, each pair of adjacent triangular surface generate a bending potential given by
\[
U_{bend} = 0.5 \cdot k_{bend} \cdot (\alpha - \alpha_{EQ})^2 \cdot \frac{l}{w}
\]
where $\alpha$ stands for the angle between both triangular surfaces, $\alpha_{EQ}$ the equilibrium angle, $l$ represents the length of the shared edge. Then tile span $w$ between both triangular surface is computed as $w=\frac{h}{6}$ with $h$ being the summed height of the triangular surfaces.

The shear potential represent the elasticity of an area of membrane and exists for each triangular surface where the particle stand as a vertex. 
\[
U_{shear} = 0.5 \cdot k_{shear} \cdot (A-A_{EQ})^2
\]
where $A$ represent the area of triangular surface and $A_{EQ}$ the area at equilibrium.

These three potentials are then summed in order to define the total elastic potential of a given particle
\[
U = U_{stretch} + U_{shear} + U_{bend}
\]
and from this total potential, the force $F_U$ on a particle $P$ is numerically derived over all three dimensions $X=(x_1, x_2, x_3)$
\[ 
F_U = \frac{dU}{dX} = \frac{U(x+\epsilon) - U(x-\epsilon)}{2 \cdot \epsilon}
\]

Another force is then added to this model in order to represent the friction caused by the membrane. This force decelerate a membrane particle by applying a negative force proportional to its velocity $v$
\[
F_{F} = -k_{friction} \cdot v
\]

Finally, Newton's law is applied to define the acceleration incurring on the particle 
\[
a =  \frac{F_U+F_F}{m} = \frac{F_U+F_F}{\tau * A} 
\]
where $\tau$ is a constant representing the shell density and $A$ the area around the particle.

In conclusion, we presented a model that has the power to simulate three dimensional elastic membranes interacting with fluid, for example the Reissner membrane and inner ear fluids. This model uses three forces to represent elasticity and takes into account the density, or thickness, of the membrane by two means : the friction force $F_{F}$ and the shell density $\tau$.

\section{Simulation of the vestibular duct}

The cochlea has a complex geometry made of multiple coiled ducts (Fig.~\ref{fig::cochlea_cut}). The cross section of each of these ducts varies from the base to the apex of the cochlea and are varying across species~\cite{Braun2012, Wysocki1999, Yost1989}. In order to better understand the various characteristics of these ducts, we start by a simplistic approximation of a duct as a simple canal and progressively make it evolve toward a more realistic cochlea duct.

\begin{figure}[t]
  \centering
  \includegraphics[width=0.86\textwidth]{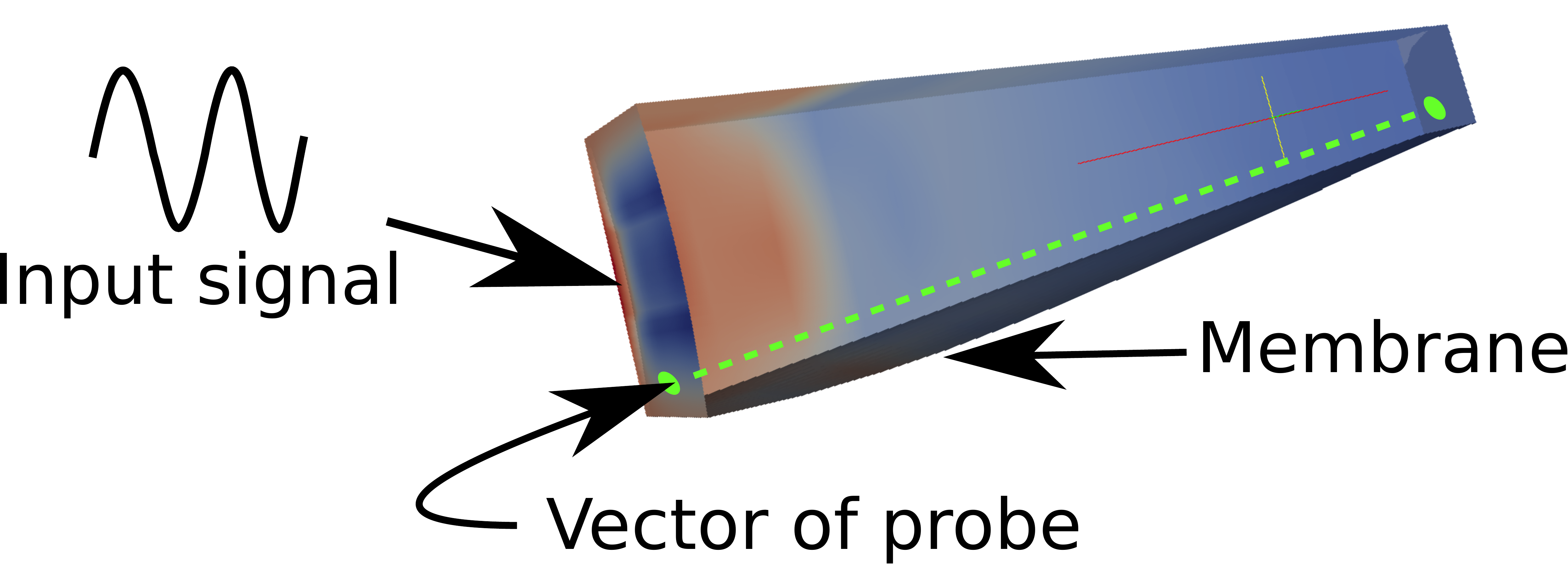}    
  \captionof{figure}{Simulation setting for the vestibular duct. Periodic vibrations are applied to the frontal lid as input signal. A vector of probe lies in vicinity of the time-dependent boundary, the membrane, in order to measure the fluid velocity over time and membrane position.}
  \label{fig:schemaSim1}
\end{figure}

We want to observe the deformation of the Reissner membrane induced by wave propagation travelling in the vestibular duct. We represented this duct as a simple rigid canal having two moving walls. One one end of the canal, the first moving wall, or membrane, is subject to periodic sinusoidal deformations (Fig.~\ref{fig:schemaSim1}). These periodic deformations imitated the mechanical vibrations of the oval window. The bottom side of the duct defined then the second membrane representing the elastic properties of the Reissner membrane. The fluid interaction on the membrane was observed using a probe vector that measured the fluid velocity close to the membrane from base to apex (Fig.~\ref{fig:schemaSim1} in green).

\subsection{Impact of the duct geometry}

Duct geometry plays a fundamental role on the propagation of mechanical waves~\cite{Kinsler1999}. Waveguides dimension and structure limit the frequency range of transported waves. In addition to these limitations, the energy of these waves decrease in function of the waveguide boundary and the medium, or fluid in our case, in which they are propagated. The fluid viscosity, heat conduction and internal molecular processes each may generate losses in the wave intensity.

In these experiments, we particularly focused on measuring the impact of the duct geometry on the velocity of travelling waves. Therefore, we did not consider a realistic setting in respect of the characteristic viscosities and frequencies of the human auditory system. We used geometries respecting the cross section to length ratio of the vestibular duct and then insured the simulation stability in order to monitor five different input vibrations (Fig.~\ref{fig:schemaSim1}). Each vibrations had its own period $w_k$ and their combination, $\sum_{k=1}^{5}\sin(w_kt)$, defined the imposed oval window deformation.

{\textbf{Simulation setting a) :}} We fixed the Reynolds number to one and the characteristic lattice velocity in lattice units to 0.01. The characteristic duct size, represented by the oval window size, was set to of $0.5$ meters. Lattice units corresponded then to $\delta x = 10^{-2}m$, $\delta t = 10^{-2}s$ and a fluid cinematic viscosity of $5\cdot10^{-3} m^2/s$ close to the one of oil. Elastic constants of the Reissner membrane ($k_{stretch}, k_{shear}, k_{bend}$) were arbitrarily fixed to $0.05$ in lattice unit. This setting required, per simulation, an average of 20 millions lattice site that were simulated over 10'000 iterations on 60 processing units for a total simulation time of one day.

\subsubsection{Rectangular or cylindrical ducts}

\begin{figure}[ht]
\center
\begin{subfigure}[t]{.32\textwidth}
  \includegraphics[width=\textwidth]{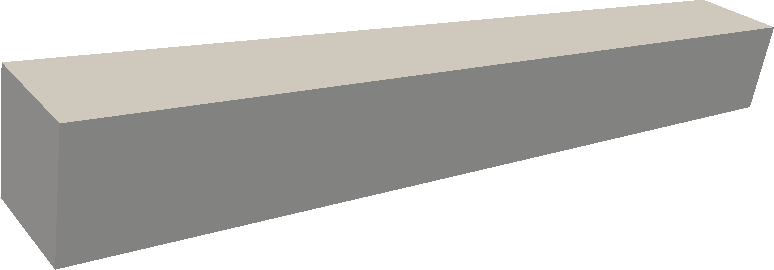}
  \caption{Cuboid}
  \label{fig:STL_Comp1_Cuboid}
\end{subfigure}
 \hfill
\begin{subfigure}[t]{.32\textwidth}
\includegraphics[width=\textwidth]{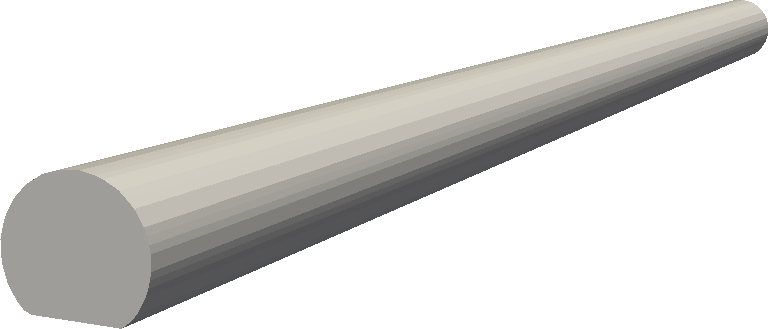}
  \caption{Cylinder}
  \label{fig:STL_Comp1_Cylinder}
\end{subfigure}
\begin{subfigure}[t]{.32\textwidth}
  \includegraphics[width=\textwidth]{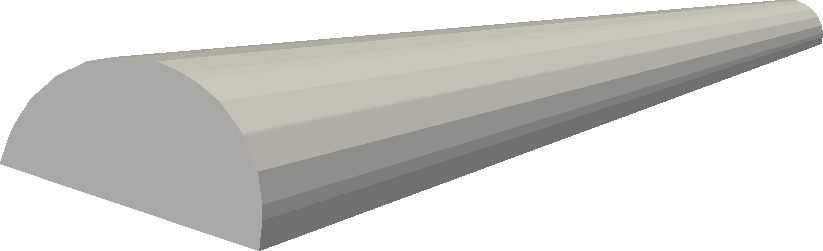}
  \caption{Half cylinder}
  \label{fig:STL_Comp1_HCylinder}
\end{subfigure}
  \caption{Geometries compared.}
  \label{fig:STL_Comp1}
\end{figure}

The aim of this first simulation is to measure the impact on wave absorption and attenuation of simplistic geometries. We are comparing three geometries represented in figure~\ref{fig:STL_Comp1} : (a) rectangular, (b) cylindrical and (c) half-cylindrical. Each of their flat bottom side is a time-dependent boundary representing the Reissner membrane.

\begin{figure}[ht]
  \centering
  \includegraphics[width=0.8\textwidth]{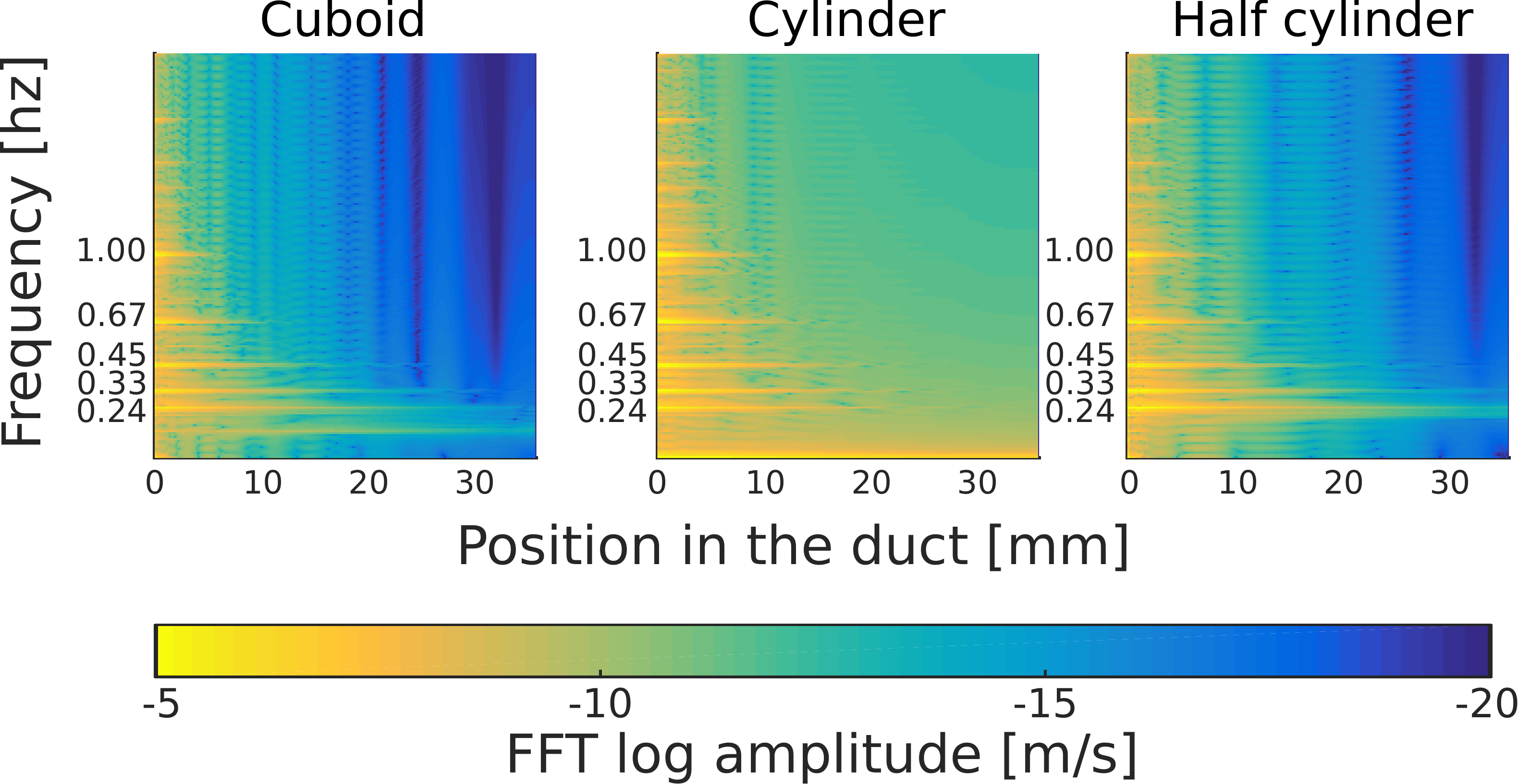}    
  \caption{Overview of the single-sided spectrum amplitude of the velocity along the membrane. X axis represent the position in the duct and Y axis the frequency. The colour indicate the value of the log single-sided spectrum amplitude of the velocity.}
  \label{fig:Comp1All}
\end{figure}

For each of these geometries, we applied vibrations at five different frequencies (0.24, 0.33, 0.45, 0.67 and 1.0 hertz) on the duct lid. We then measured the fluid velocity close to the membrane and thus obtained temporal measures of velocity for 1200 points along the membrane. For each spatial points, we computed the frequency spectrum of the fluid velocity over time using a discrete Fourier transform. The single-sided amplitude spectrum\footnote{The single-sided amplitude spectrum represents the sum of the absolute amplitude of negative and positive frequencies.} was then extracted and represented against the position in the membrane and the frequency (Fig.~\ref{fig:Comp1All}).

The five input frequencies are well identifiable for all three geometries. In all of them, the velocity of waves with higher frequencies decrease faster along the membrane than waves with low frequencies. For each of the imposed vibrations, multiple harmonics appear. However, their weaker velocity decay in the first third of the duct.

This overview shows that these three geometries propagate waves differently. While the same waves frequency are observed, their amplitude along the membrane differ. We postulate that the cuboid absorb more wave energy than the cylinder because of the amount of flat surface that reflect waves. Inded such reflection may cause interferences that attenuate the wave energy. 

\begin{figure}[ht]
  \centering
  \includegraphics[width=0.8\textwidth]{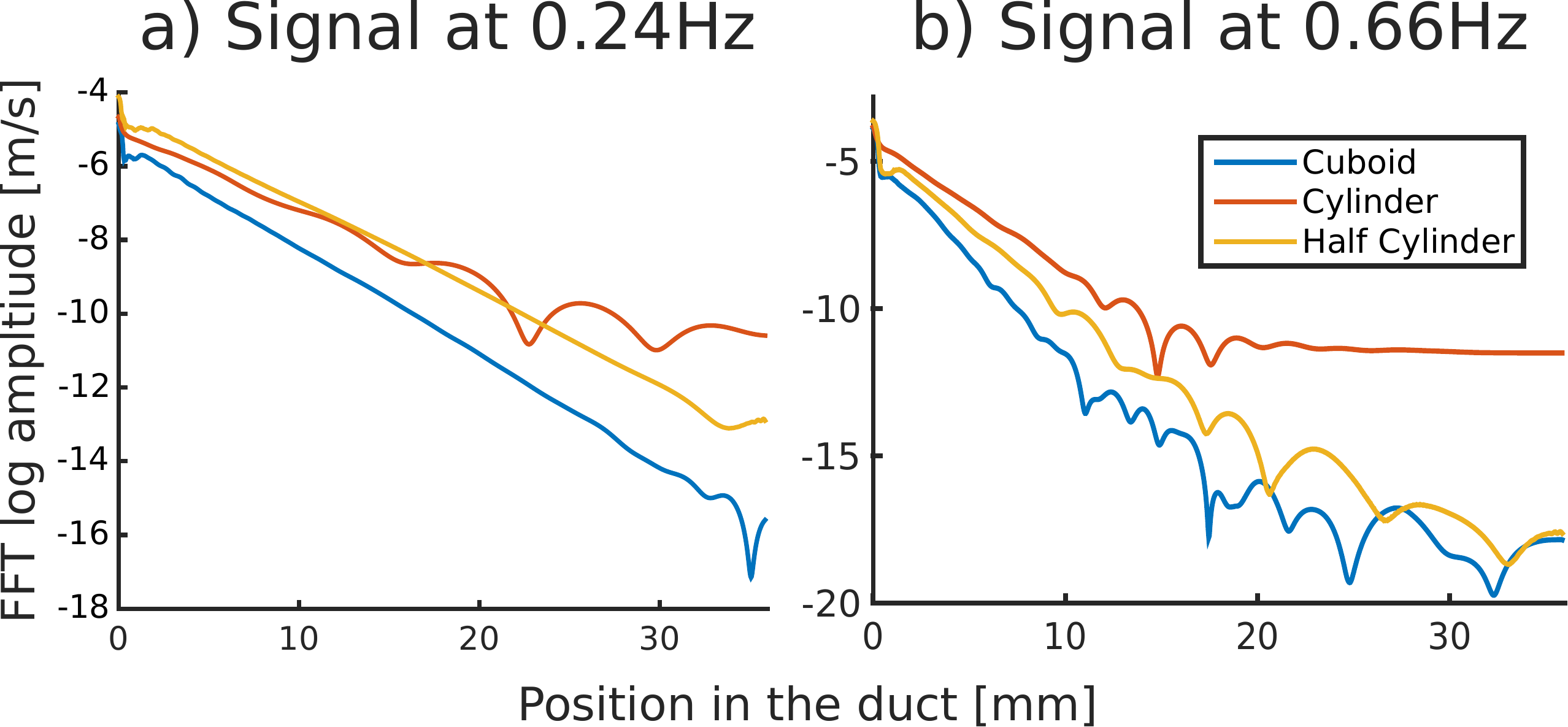}    
  \caption{Comparison of single-sided velocity spectrum amplitude at isolated frequencies : 0.24 and 0.66 hertz.}
  \label{fig:Comp1Freq}
\end{figure}

At the imposed 0.24 and 0.66 hertz input frequencies (Fig.~\ref{fig:Comp1Freq}), waves in the cylinder are more stably propagated than in the cuboid. The half cylinder mixes the structural properties of both other geometries and thus lies in between them in term of wave propagation. While the absorption of wave velocity along the membrane differs for each geometries, it doesn't impact the key phenomena that we are observing : high frequencies waves dissipate faster than low frequencies ones along the duct.

\subsubsection{Structural characteristics of the vestibular duct}
\label{sec:structural}
\begin{figure}[ht]
  \centering
  \includegraphics[width=0.8\textwidth]{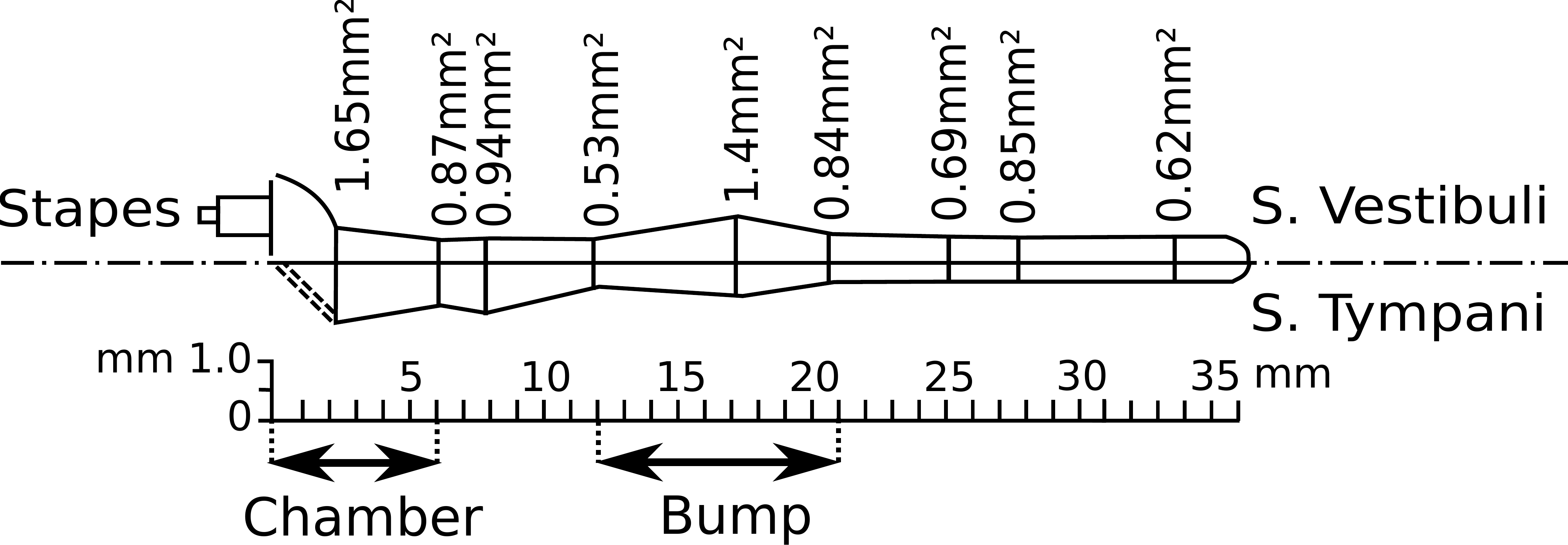}    
  \caption{Cross section dimension of the scala vestibuli and main feature of its structure : chamber and bump. Schema and measures adapted from~\cite{Braun2012, Wysocki1999, Yost1989}.}
  \label{fig:ScalaDim}
\end{figure}

Cross section of cochlea ducts does not linearly decrease from base to apex. In the human inner ear, the cross section of the vestibular duct varies frequently. Close to its base, a large chamber progressively tighten over approximately $6$ millimetres. It then reaches a thin plateau until a bump starts and reaches its end $10$ millimetres farther (Fig.~\ref{fig:ScalaDim}).

Such changes of cross section in a waveguide influence the propagated waves differently given their frequency~\cite{Kinsler1999}. Indeed, frequencies can be filtered in function of widening or tightening of the the duct. When a duct grow in size over a given distance, it acts as a low-pass acoustic filter. A low-pass filter propagate waves with frequencies lower than its cut-off frequency while attenuating higher frequencies. A high-pass acoustic filter is the opposite, it occurs when the duct shrink over a given distance and passes frequencies higher than its cut-off frequency.

\begin{figure}[ht]
  \centering
\begin{subfigure}{.45\textwidth}
  \includegraphics[width=\textwidth]{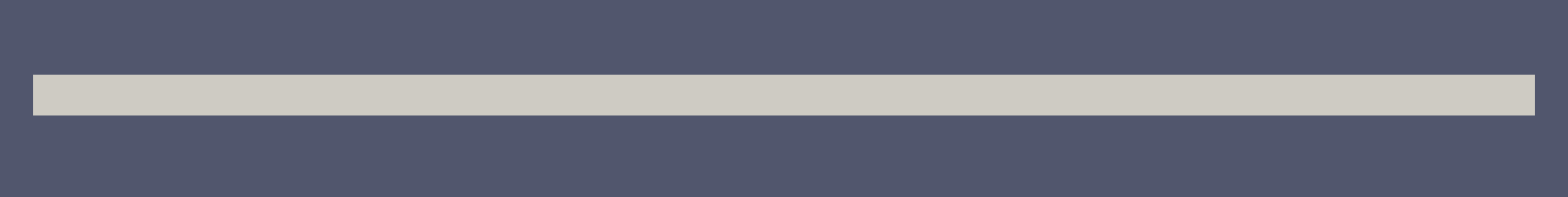}
  \caption{\textbf{Cuboid}}
  %\label{fig:STL_Comp2_Cuboid}
\end{subfigure}
\hfill
\begin{subfigure}{.45\textwidth}
  \includegraphics[width=\textwidth]{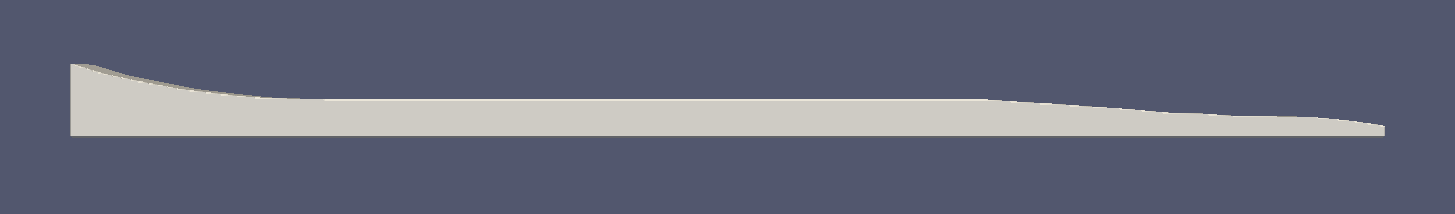}
  \captionof{figure}{With \textbf{chamber}}
  %\label{fig:STL_Comp2_Chamber}
\end{subfigure}
\\
\begin{subfigure}{.45\textwidth}
\includegraphics[width=\textwidth]{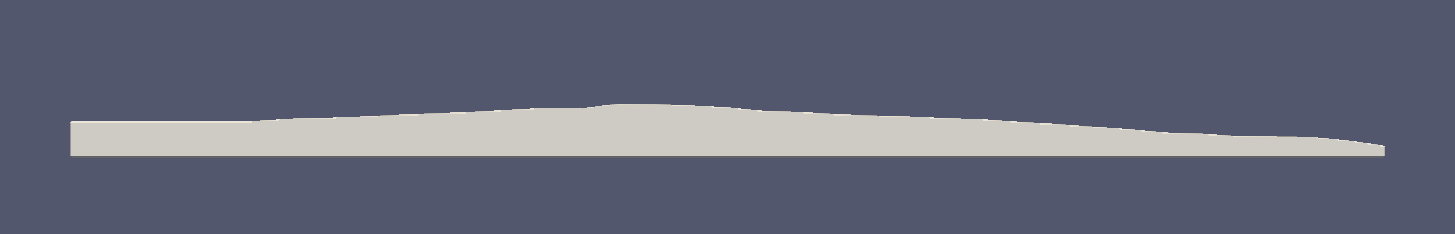}
  \caption{Without \textbf{bump}}
  %\label{fig:STL_Comp2_Bump}
\end{subfigure}
\hfill
\begin{subfigure}{.45\textwidth}
  \includegraphics[width=\textwidth]{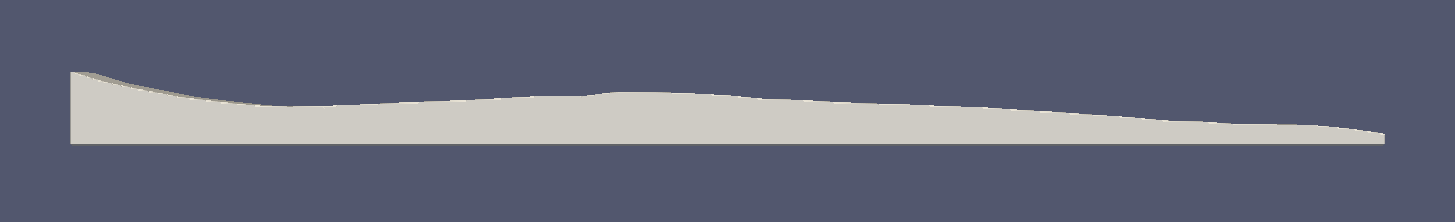}
  \caption{With \textbf{both} chamber and bump}
  %\label{fig:STL_Comp2_Both}
\end{subfigure}
  \caption{Geometries compared in order to measures the frequency filtering effect of their structure.}
  \label{fig:STL_Comp2}
\end{figure}

The observed characteristics of the vestibular duct let support the presence of such filtering effects. Therefore, we compared four different geometries (Fig.~\ref{fig:STL_Comp2}) in order to monitor if such filtering take place. We used the (a) cuboid geometry as reference having no filtering effects. Its cross section stands constant over all the duct length. The two main features of the vestibular duct are then evaluated separately by modifying the base geometry. The first one (b) is the widening at the base of the duct that form a chamber over $6$ millimetres. The second one (c) is the widening in the middle of the duct that form a bump over $10$ millimetres. Finally, we combined both feature (d) in order to observe the potential joined filtering effects.

\begin{figure}[ht]
  \centering
  \includegraphics[width=0.9\textwidth]{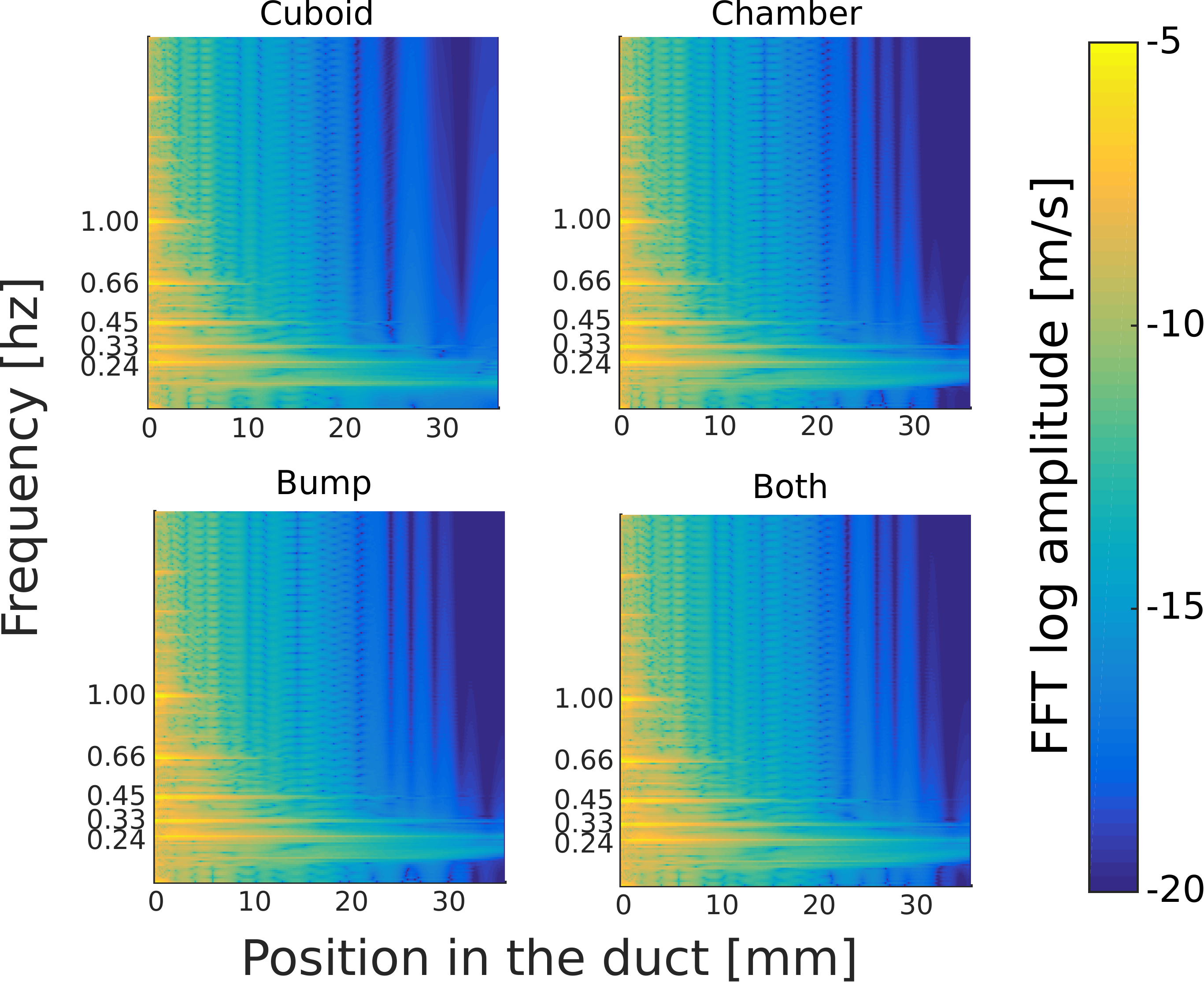}    
  \caption{Overview of the single-sided spectrum amplitude of the velocity along the membrane.}
  \label{fig:Comp2_All}
\end{figure}
 
The measuring setting was the same as in the previous simulations : the same five frequencies were imposed on the lid of each geometries and the velocity was measured along the duct. We extracted from these measures the velocity single sided spectrum amplitude along the membrane using a discrete Fourier transform. Figure~\ref{fig:Comp2_All} gives an overview of the velocity spectrum amplitude against the frequency and duct location. Once again, the five input frequencies are well propagated on each geometry.

\begin{figure}[ht]
  \centering
  \includegraphics[width=0.7\textwidth]{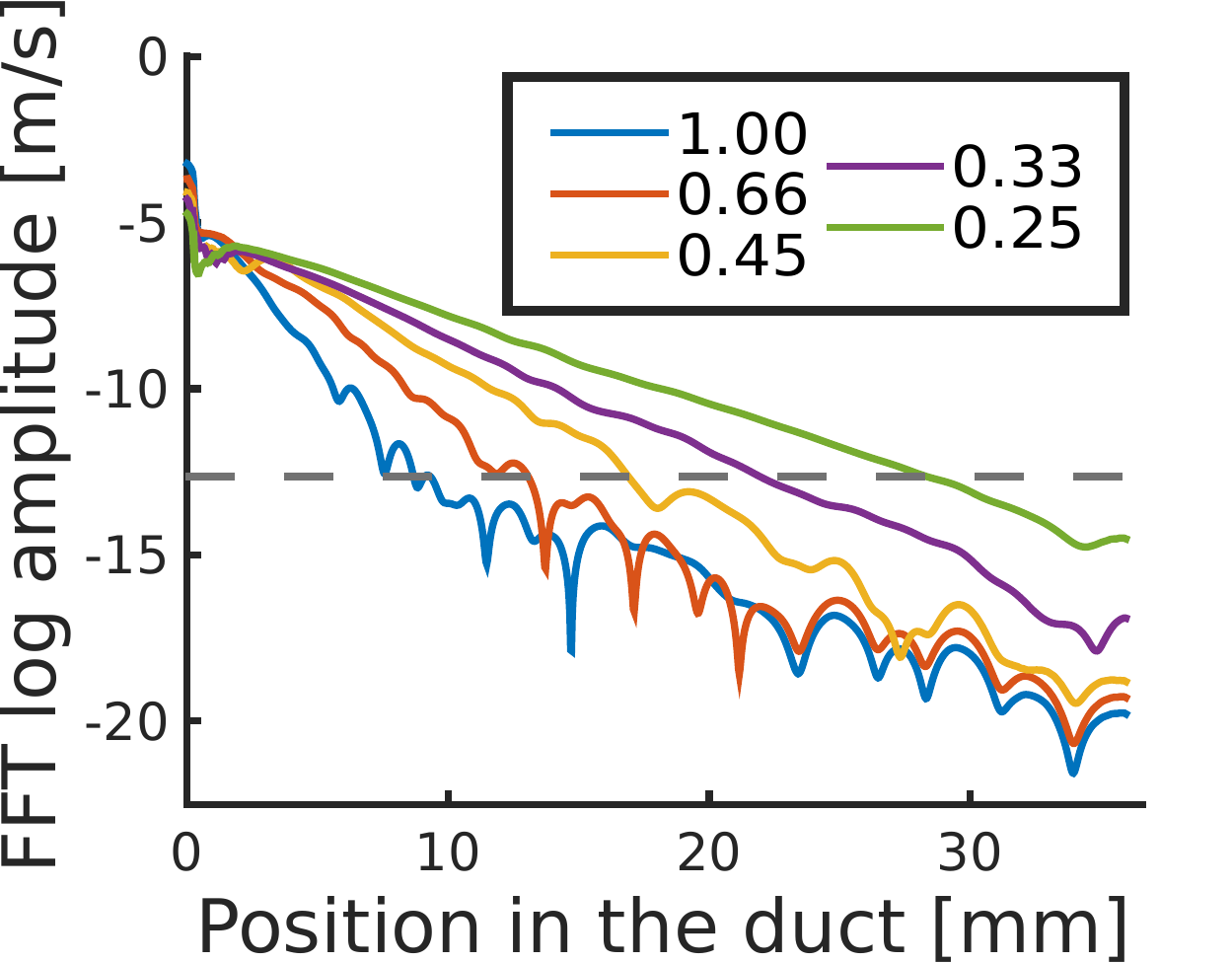}
  \caption{Single-sided spectrum amplitude of the velocity along the membrane at input frequencies for the geometry with both features. The grey dotted line represents the observed noise threshold on high frequencies signal.}
  \label{fig:Comp2_Freq_SVRatio}
\end{figure}

Figure~\ref{fig:Comp2_Freq_SVRatio} details more accurately the velocity amplitude of each input frequency in the geometry (c). These measures shows that once again the decay of velocity amplitude is quicker for the \textit{higher} frequencies. In addition, this figure shows that the signal becomes noisy and unstable when the velocity amplitude reaches $10^{-12.5}m/s$.

In order to quantify these filtering effects, we compare the velocity amplitude at 0.24 and 0.66 hertz of modified geometries (b,c,d) with the one of the base rectangular geometry (a). The amplifications are showed for amplitude greater than $10^{-12.5}[m/s]$. Under this value noise appears and disrupts the measured velocity amplitude.

\begin{figure}[ht]
\centering
\includegraphics[width=0.9\textwidth]{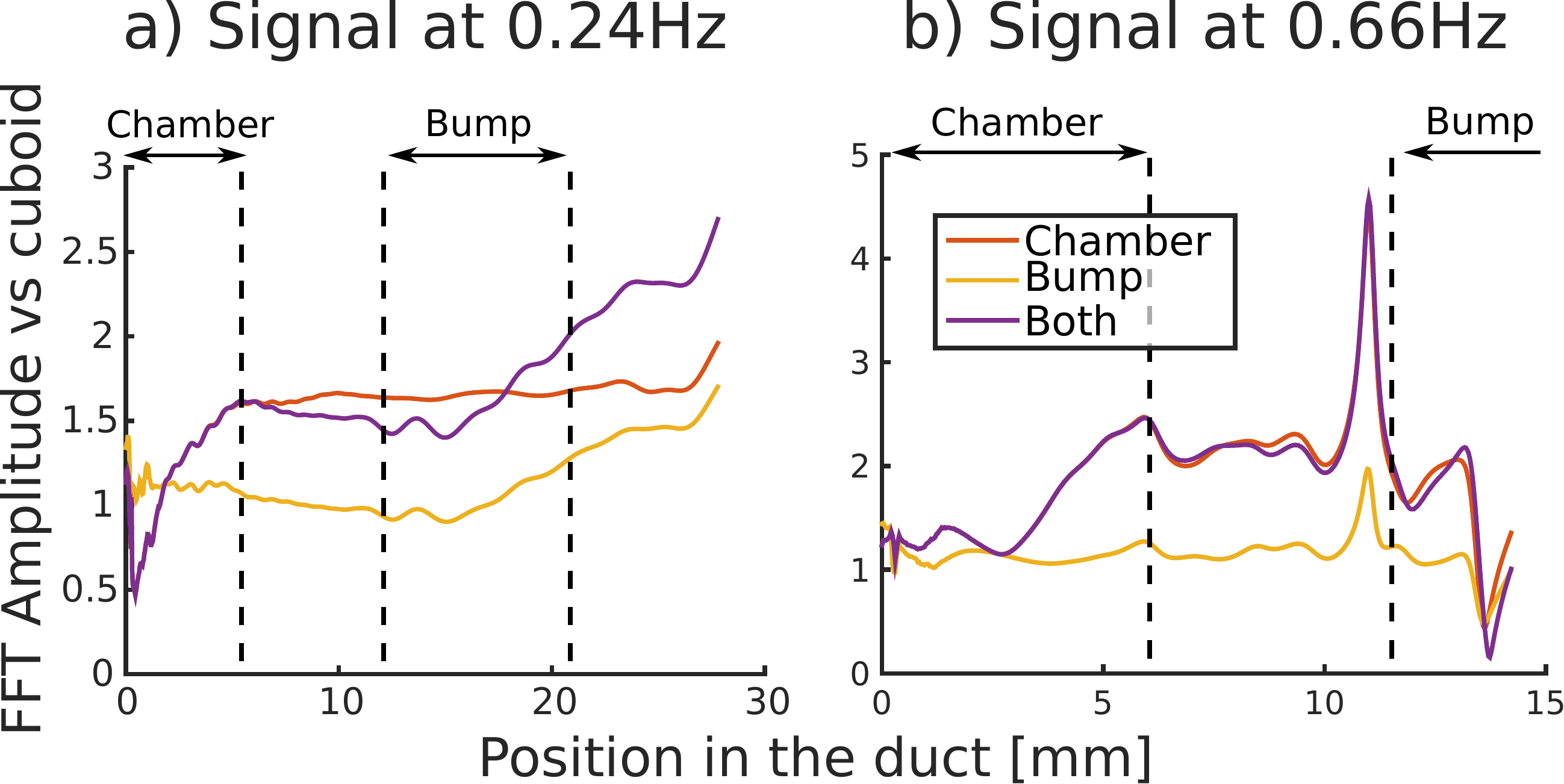}
\caption{Amplification of the velocity spectrum amplitude of modified geometries when compared to the cuboid.}
\label{fig:Comp2_Ampl_Both}  
\end{figure}

Figure.~\ref{fig:Comp2_Ampl_Both} reveals that the chamber has a considerable impact on waves at both frequencies. It amplifies them by at least a twofold factor. In between the chamber and the bump, lies a small plateau that seems to act as a high-pass filter. The low frequency signal is slightly attenuated in this region while the high frequency signal remains untouched. The effect of the bump is relatively more difficult to quantify. While it seems that it amplify the low frequency signal, it has a rather unsettling effect on the high frequency signal. Given that in the bump vicinity, the measured velocity amplitude is subject to noise, both effects are hard to assess with confidence. 

From these measures we conclude that the structure of the vestibular duct filters wave in function of their frequencies. The chamber seems to amplify signals, more so high frequencies one. The small plateau following seems to maintain the frequencies of high frequencies waves while attenuating lower one. The bump probably amplify wave at low frequencies. Our analyses being based on the velocity norm, we hypothesize that a finer study of each separate velocity components would help to better understand the effect of these duct cavity.

\clearpage

% Comparison of STLs, impact of geometry on measured
\subsection{Emulating the human cochlea conditions}
\label{sec:HumanEnv}

In our previous simulations, we were focused on analysing the vestibular duct structural effects on the propagated waves. In order to ensure the stability of these simulations, we used physical properties differing strongly from the one of mammals. In this section, we aim at increasing the realism of our simulation by using the physical properties of the human ear. In order to achieve such objectives, two points have to be addressed. In a first time, we have to insure stable simulations based on the properties of the human cochlea. Then, in a second time, we have to directly measure the deformation of the Reissner membrane.

{\textbf{Simulation setting b) :}} In order to fulfil the first point, we considered the viscosity of the perilymph and the endolymph. These fluids cinematic viscosity is of $8\cdot10^{-7} m^2/s$ that correspond to mineralized water~\cite{Yost1989}. The characteristic dimension of the input membrane was set to $10^{-3}$ meters with a duct length of $36\cdot10^{-3}m$. Lattice units corresponded then to $\delta x = 2\cdot10^{-5}m$, $\delta t = 10^{-5}s$ and a Reynolds value of $125$. This setting required, per simulation, an average of 20 millions lattice sites that were simulated over 20'000 iterations on 48 processing units for a total simulation time of one day.

In the following experiments, we only considered one frequency at a time. Such choice was made to stabilize the simulation and reduce the total runtime for each of them. Indeed, simulating two frequencies separated by a hundredfold factor, i.e. 50 and 5000 hertz, require a considerable computational effort. This comes from the need to guarantee a sufficient sampling rate for the high frequency wave by decreasing $\delta t$ and to generate enough iterations to measure multiple period of the low frequency wave. A gross estimation of the time increase gives that covering 50 and 5000 hertz in the same simulation would then take 100 times longer than just one of them.

We arbitrarily chose an input vibration frequency of 100 hertz. This frequency produced stable simulations and solicited fluid movement in a large portion of the duct. While being less stable, simulations with input vibrations at higher frequencies showed similar phenomenons and maintained the ones observed in figures~2.5 and 2.6 : high frequencies waves dissipated faster than low frequencies ones along the duct.

\subsubsection{Attenuation and absorption of waves}

Our simulation setting, illustrated by figures~\ref{fig:schemaSim1}, reveals that the external side of the membrane is in contact with air. However in the inner ear, the cochlear duct sits at the other side of the Reissner membrane. Palabos doesn't yet offer to simulate immersed time-dependent elastic boundary. This is problematic since moving a membrane immersed by salty water on both side does not equate to moving a membrane with air on one side. 

In addition to model inaccuracies, we have to deal with multiple other knowledge gaps such as the lack of data on the Reissner membrane mass and elastic properties. Therefore, we postulate that replacing the air present on one side of the membrane by a fluid would absorb kinetic energy and thus slows the membrane displacement. We accordingly tune the shell density and the friction force in order to absorb waves energy and to produce simulation settings closer to the real environment of the human cochlea.

\begin{figure}[ht]
    \centering
\includegraphics[width=0.9\textwidth]{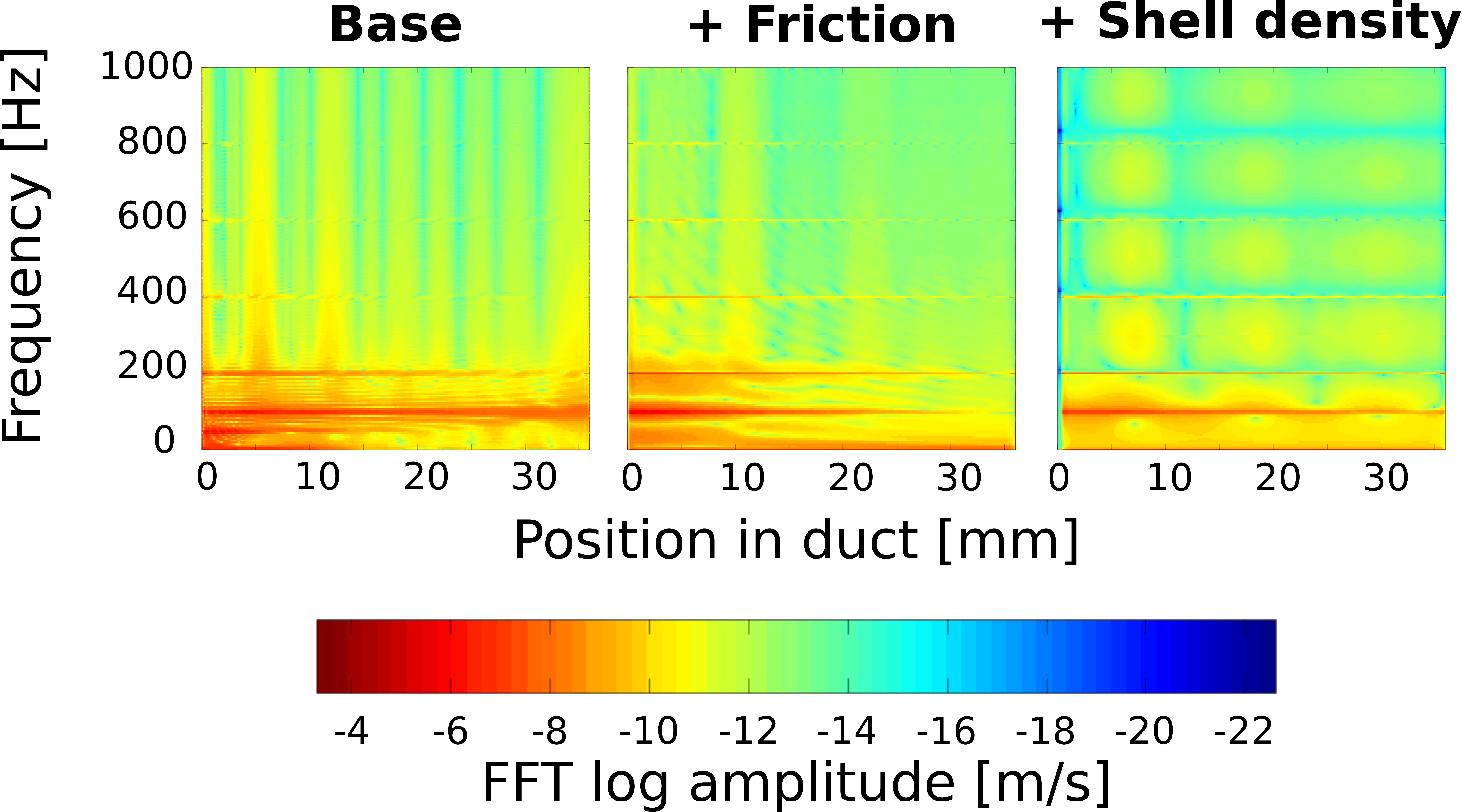}
  \caption{Comparison of the impact of the Reissner membrane shell density and friction force parameters when subject to an 100 hertz input vibration on the oval window. The single-sided velocity spectrum amplitude is showed with, from left to right : 1) The base model without friction and low shell density; 2) Addition of membrane friction; 3) Increase of shell density.}
\label{fig:improve}
\end{figure}

As in previous experiments, we extract the one-sided spectral amplitude of the velocity obtained from the probes vector lying close to the membrane (Fig.~\ref{fig:improve}). The base simulation correspond to the \textit{simulation setting b)} defined previously and an input vibration at 100 hertz. With this setting, the amplitude at 100 hertz is so strong that it goes all the way trough the duct and bounce on the opposite lid. Moreover, the lack of energy absorption creates a strong harmonic wave that appears at 200 hertz. In order, to reduce this phenomena we added a friction coefficient to the membrane. This friction reduced strongly the amplitude of the velocity but amplified the harmonics of the main frequency. Increasing the shell density hindered this effect and produced a clear signal at 100 hertz with weaker harmonics.

\subsubsection{Measuring Reissner membrane deformation}

While the velocity of the fluid offers a good indication of the Reissner membrane movement, we need a more accurate measure of its deformation in order to link both the vestibular and the cochlear duct. Indeed, our experiment goal is to monitor the effect that a vibration on the lid of the vestibular duct would induce in the microscopic organ of Corti resting in the cochlear duct. Linking accurately both duct is therefore of utmost importance. 

In this experiment, we measure the physical displacement of the membrane. For that, we have to remember that this membrane is represented by a set of membrane particles that \textit{freely} evolve in a three dimensional space. More precisely, tenth of thousands of particles are required to correctly represent the membrane. Tracking such amount of particles over tenth of thousand of iterations is a daunting task : it would require a large amount of memory. 

\begin{figure}[ht]
  \center
  \includegraphics[width=0.9\textwidth]{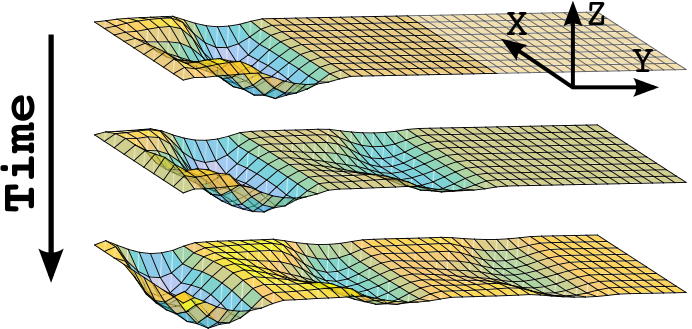}
  \caption{Measured deformation of the membrane at various time steps. Colours represents the height of the membrane going from blue to yellow to red. \label{fig:waves}}
  
\end{figure}

In order to reduce the cost of measuring the membrane deformation, we used the two-dimensional (X- and Y-axis) flat plane formed by the membrane at rest as reference. We then reduced the number of points required to memorize the deformation by using a discrete representation of this plane, with a coarse resolution when compared to the number of membrane particles. Instead of tracking all particles, we measured the membrane deformation only in height (Z-axis) at each fix points (X- and Y-axis) of this discrete plane. Figure~\ref{fig:waves} illustrate deformations measured at different time steps.

\begin{figure}[ht]

  \center
  \begin{subfigure}[t]{.45\textwidth}
  \includegraphics[width=\textwidth]{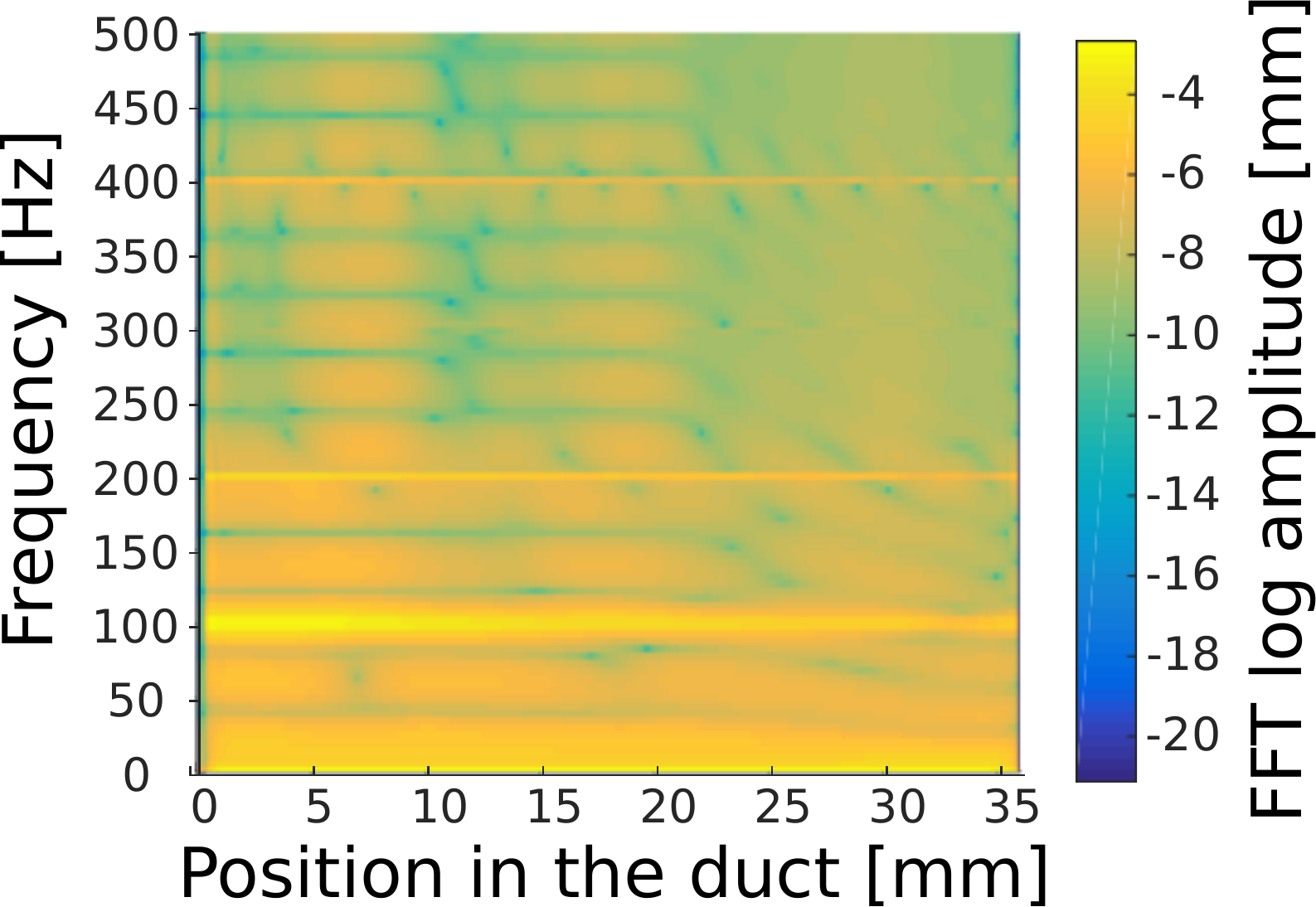}
  \caption{Overview.}
  \label{fig:MembraneDeformAll}

  \end{subfigure}
  \hfill
  \begin{subfigure}[t]{.45\textwidth}
  \includegraphics[width=\textwidth]{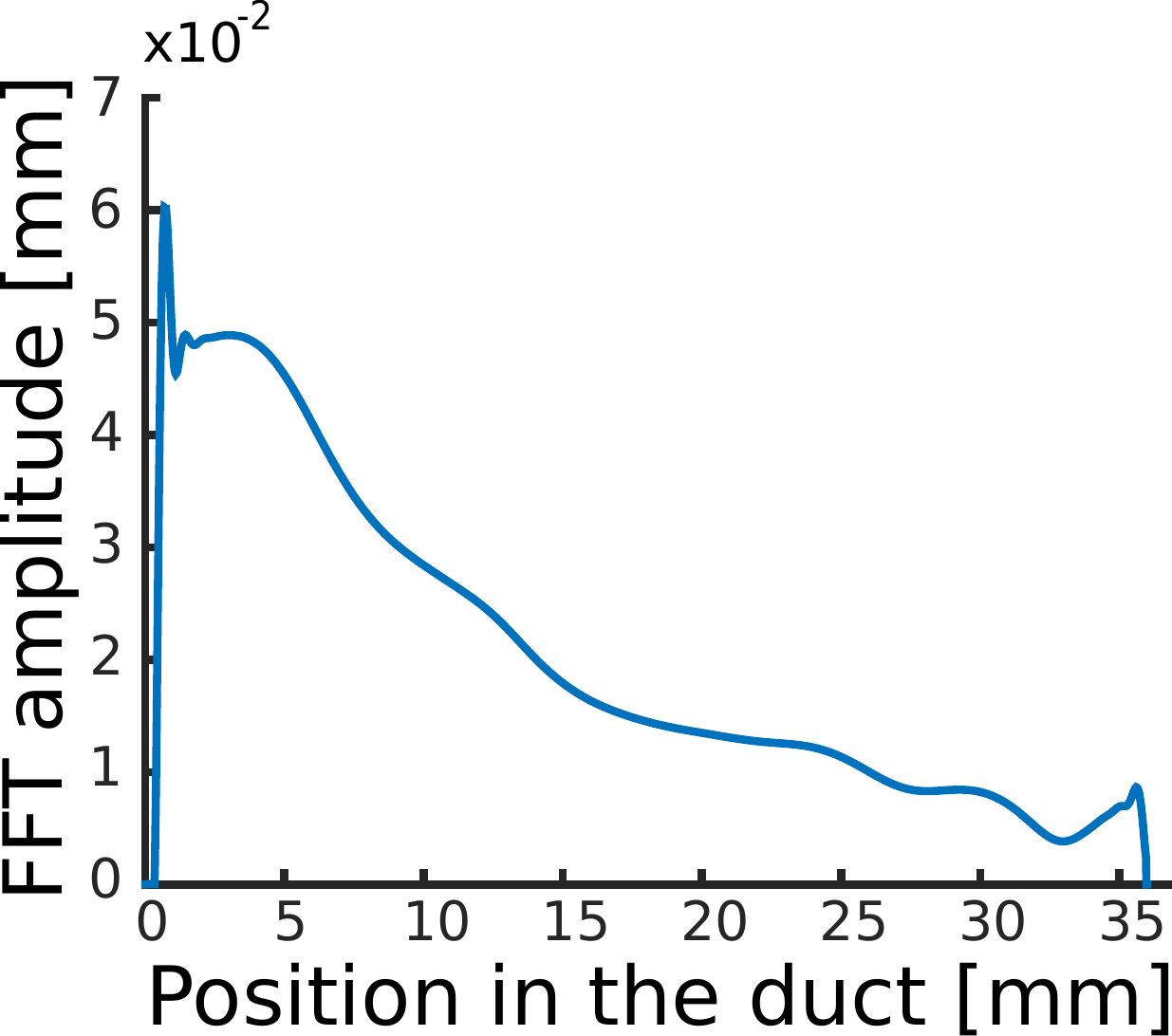}
  \caption{Isolated at 100 hertz. }
  \label{fig:MembraneDeform100Hz}
  \end{subfigure}
  
     \caption{Single-sided amplitude spectrum of the membrane deformation. }  
     \label{fig:membraneDeform}

\end{figure}

Using the previously defined realistic conditions, we memorized the membrane deformation in response to an input vibration at 100 hertz. The choice of this frequency was made in order to observe a deformation all along the membrane. We extracted the central vector of deformation measures in respect of the X-axis. We choose this measures vector as it represented the biggest observed membrane deformation over time and analysed its one-sided spectrum amplitude. 

Figure~\ref{fig:MembraneDeformAll} shows an oscillation of the membrane at 100 hertz with weaker harmonics at 200 and 400 hertz. These observations corroborate the one previously made on the fluid velocity. Figure~\ref{fig:MembraneDeformAll} isolates the membrane displacement amplitude for the 100 hertz frequency. The amplitude starts with a value of 50 microns and reduces until the end of the membrane where it reaches a value of 10 microns. The order of magnitude of this simulated deformation is in the same range as the measured thickness of the Reissner membrane (5-15 microns~\cite{Shibata2009}). 

It is rather interesting to note that this amplitude begins to drop after the structural chamber of the vestibular duct. In section~\ref{sec:structural}, we stipulated that the structural plateau in this region may act as a high-pass filter. And indeed, the major part of the deformation attenuation happens in between $6$ and $11-15$ millimetres which roughly correspond to the end of the chamber and the start of the bump. 

\section[Simulation of the cochlear duct and the organ of Corti]{Simulation of the cochlear duct and the organ of Corti\sectionmark{Cochlear duct and Corti's organ}}
\sectionmark{Cochlear duct and Corti's organ}

% Type of STL used
% What we want to measure
In the previous section, we have shown that periodic waves propagating in the vestibular duct produce oscillation of the Reissner membrane of equal frequencies. This membrane connects the vestibular duct to the cochlear duct. Therefore, this oscillation may well create fluid movements in the cochlear duct and propagate to the microscopic organ of Corti. In this section, we enquire if such phenomena could happen and explain how this organ micro-mechanics translates mechanical waves into nervous signals.

\begin{figure}[ht]
  \center
  \includegraphics[width=\textwidth]{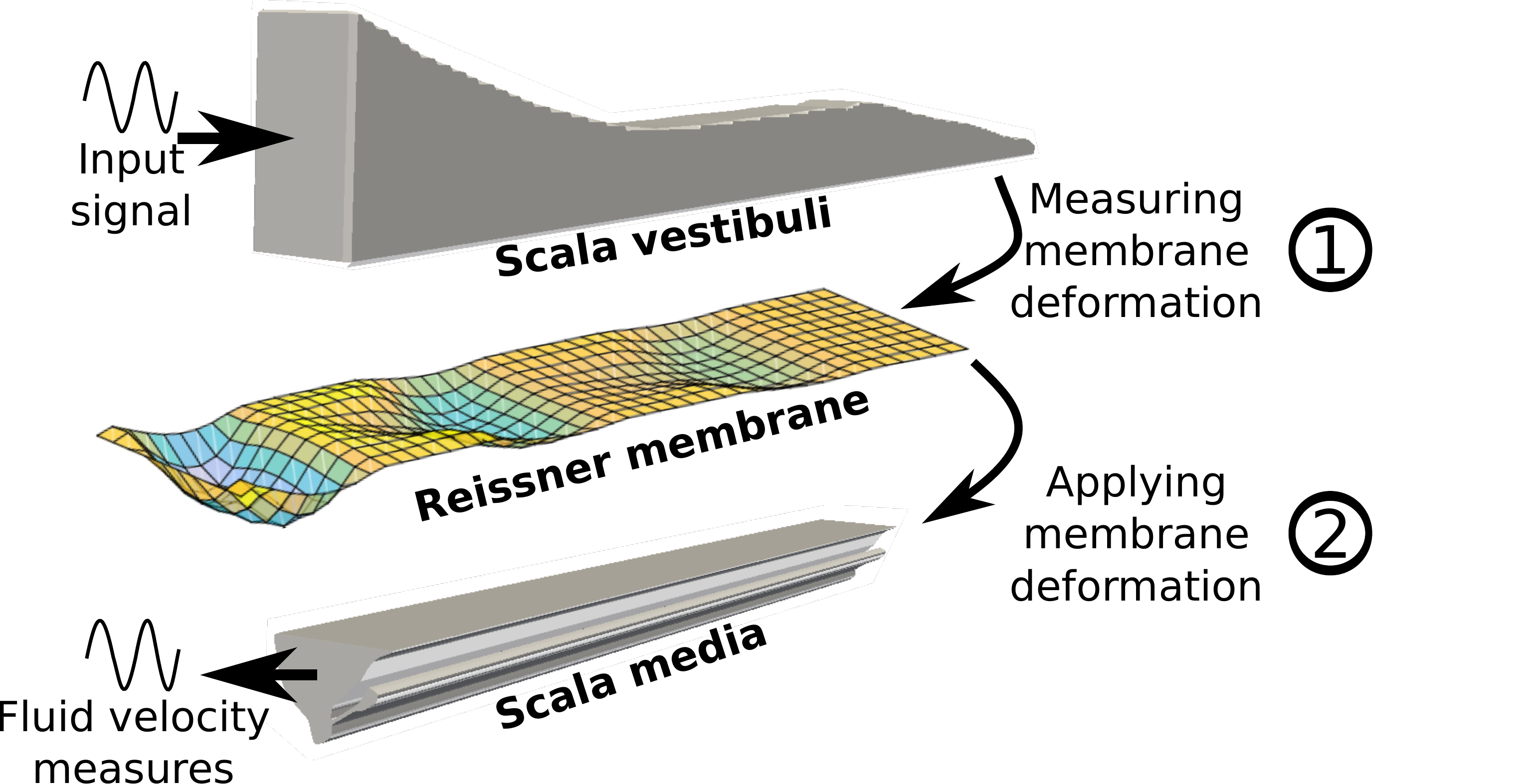}
  \caption{Simulation setting for the scala media. Deformation measure obtained during the first phase are applied to the Reissner membrane. Vectors of probes are placed at key points in the duct in order to measures temporal and spatial fluid velocity.}
  \label{fig:ScalaMediaSettings}
\end{figure}

Figure~\ref{fig:ScalaMediaSettings} shows the two main steps of this experiment. The first one correspond to the measures of the Reissner membrane deformation with the \textit{simulation setting b)} described in section~\ref{sec:HumanEnv}. In the second step, we applied this deformation on the the geometry representing the cochlear duct. 

\begin{figure}[ht]
  \center
  \includegraphics[width=\textwidth]{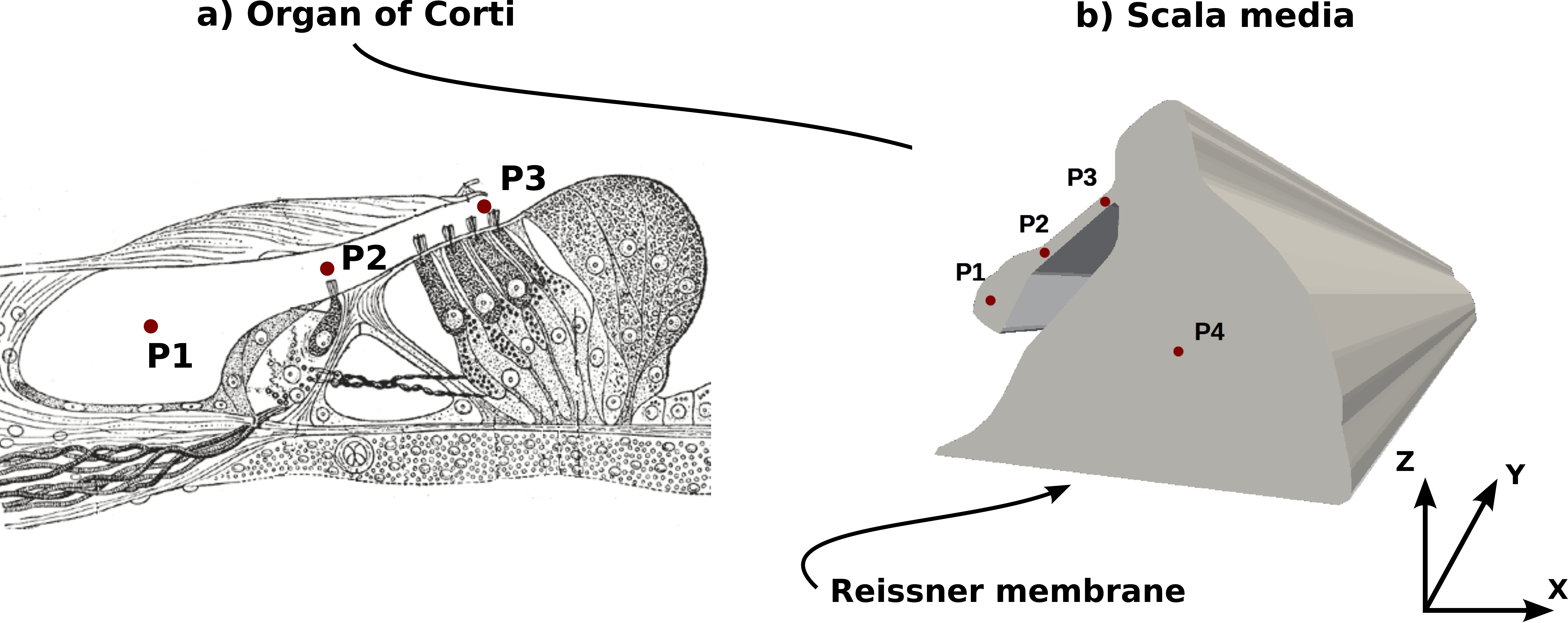}
  \caption{Figure a) shows the biological representation of the organ of Corti and the placement of the probes (adapted from~\cite{Gray1918}). Figure b) shows the scala media, horizontally flipped with respect to its representation in figure~3.1, and the probes positions.}
  \label{fig:ScalaMediaProbes}
\end{figure}

However, this geometry, represented in figure~\ref{fig:ScalaMediaProbes}, is far more complex than the one of the vestibular duct and thus was designed more accurately. Such level of detail was required to correctly simulated the fluid velocity inside the organ of Corti. Therefore this canal proportion and global aspect were carefully preserved during its conception. The points of interest lies in the organ of Corti (P1), the scala media (P4) and the tiny canal linking both elements (P2, P3). Fluid velocity of each of these locations will therefore be measured.

\subsection{Coupling both simulations}

In order to guarantee the coupling of simulation, as shown in figure~\ref{fig:ScalaMediaSettings}, the plane formed by the Reissner membrane was used as a reference. However, while the vestibular duct geometry is long of approximately 35 millimetres, as in reality, we modelled the cochlear duct with a length of one third its true size ($10$ millimetres). This choice was made to reduce the computing complexity of the simulations. 

The number of lattice sites in the simulation domain is function of the geometry dimension and the simulation spatial resolution. This resolution is conditioned by the smallest structure in the geometry. In this case, we speak of the entrance of Corti's organ (P2, P3) with a cross sectional size of less than 20 microns. With the duct length of 35 millimetres as the biggest dimension, the size ratio between the tiny canal and the duct length would be greater than 1500. Therefore, we chose a length of $10$ millimetres in order to correctly observe the velocity in the first third cochlear duct. Length that already produce a computationally challenging simulation having a size ratio of 500 between its smaller and bigger elements.

{\textbf{Simulation setting c) :}} The \textit{simulation setting b)} described in section~\ref{sec:HumanEnv} were used for this simulation. The spatial resolution of the second simulation was however increased in order to adequately represent the details of Corti's organ ($\delta x = 10^{-6}m$). One third of the cochlear duct required 6 millions of lattice sites to be represented which were simulated over 20'000 iterations on 96 processing cores for a total simulation time of one day. This duct complex geometry explains the increase of computing power when compared to simulation of the vestibular duct. Indeed, computation of fluid interactions with boundary layers cost more than the computation of inner fluid sites.

\subsection{Velocity of the fluid in the scala media}

We used four probes in the middle of the duct (middle of Y-axis) represented on figure~\ref{fig:ScalaMediaProbes} : $P1, P2, P2, P4$. The fluid velocity in the duct was measured at each of those points. For this simulation, the separate velocity components along the X, Y and Z-axis were recorded in order accurately analyse the direction of the fluid. The single-sided velocity amplitude spectrum at each of these point was computed by discrete Fourier transform. 

\begin{figure}[ht]
    \centering
\begin{subfigure}{.42\textwidth}
  \includegraphics[width=\textwidth]{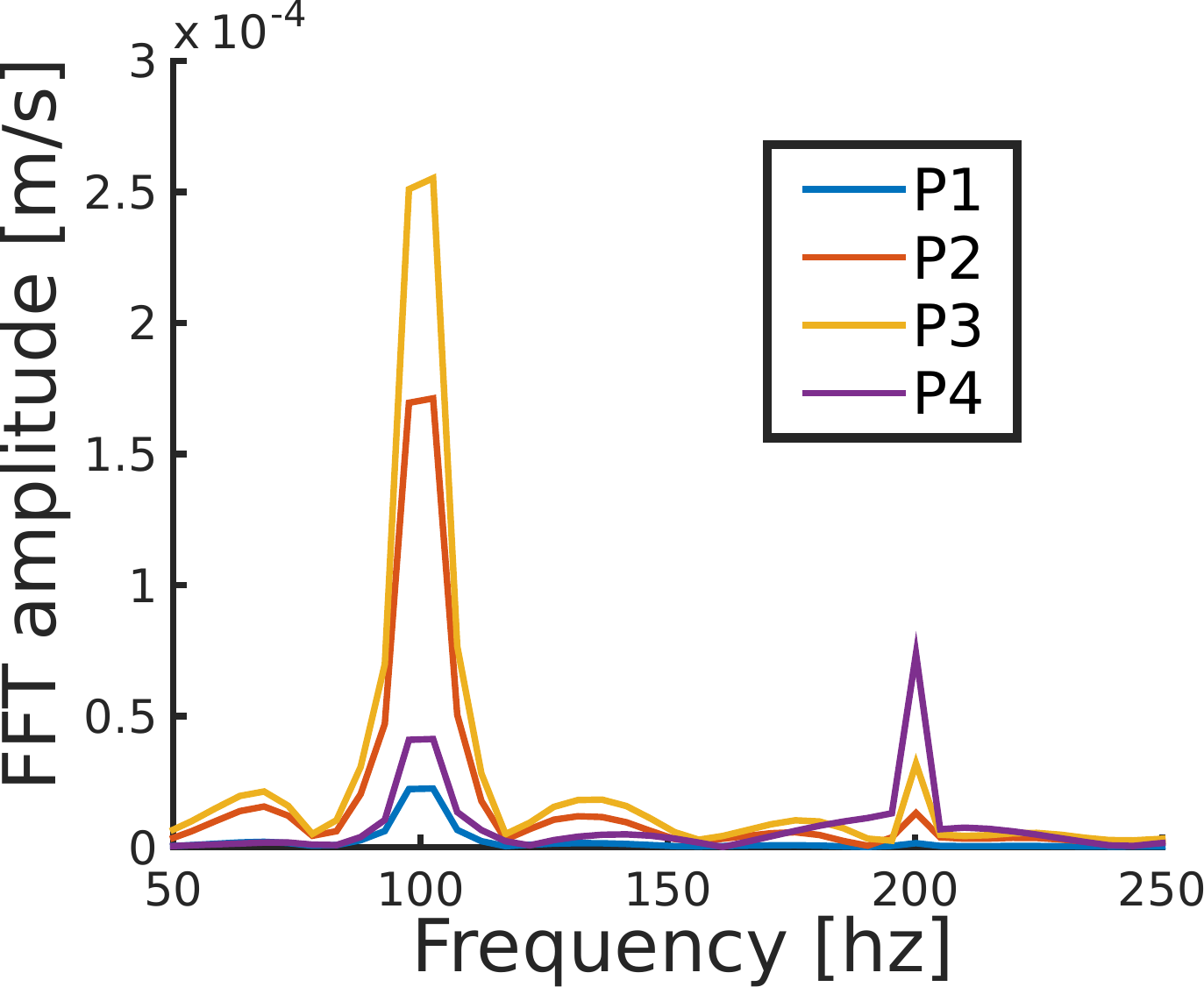}
  \caption{X axis velocity component.}
  \label{fig:ResSM_Vx}
\end{subfigure}
\hfill
\begin{subfigure}{.42\textwidth}
  \includegraphics[width=\textwidth]{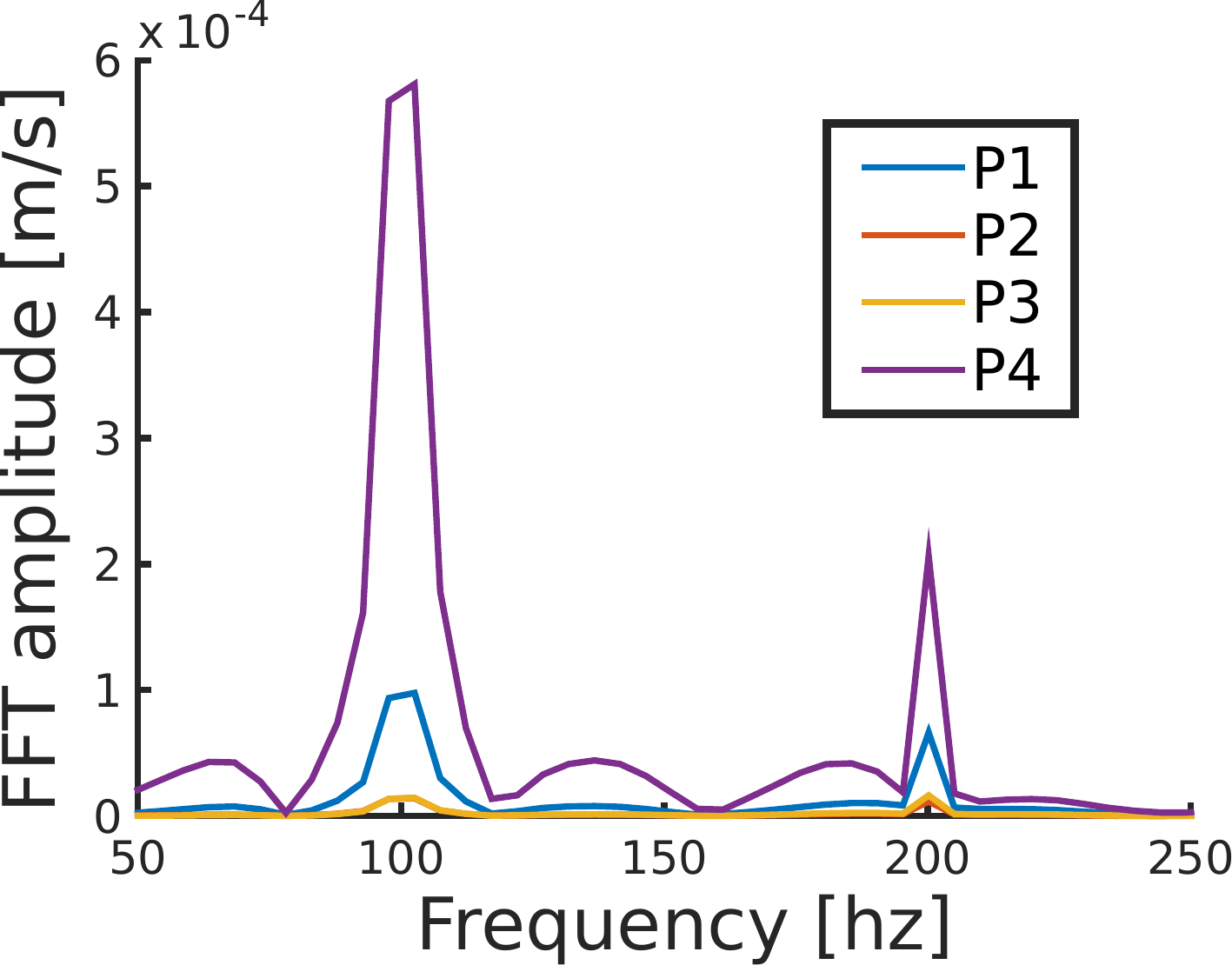}
  \caption{Y axis velocity component.}
  \label{fig:ResSM_Vy}
\end{subfigure}
\\
\begin{subfigure}{.42\textwidth}
  \includegraphics[width=\textwidth]{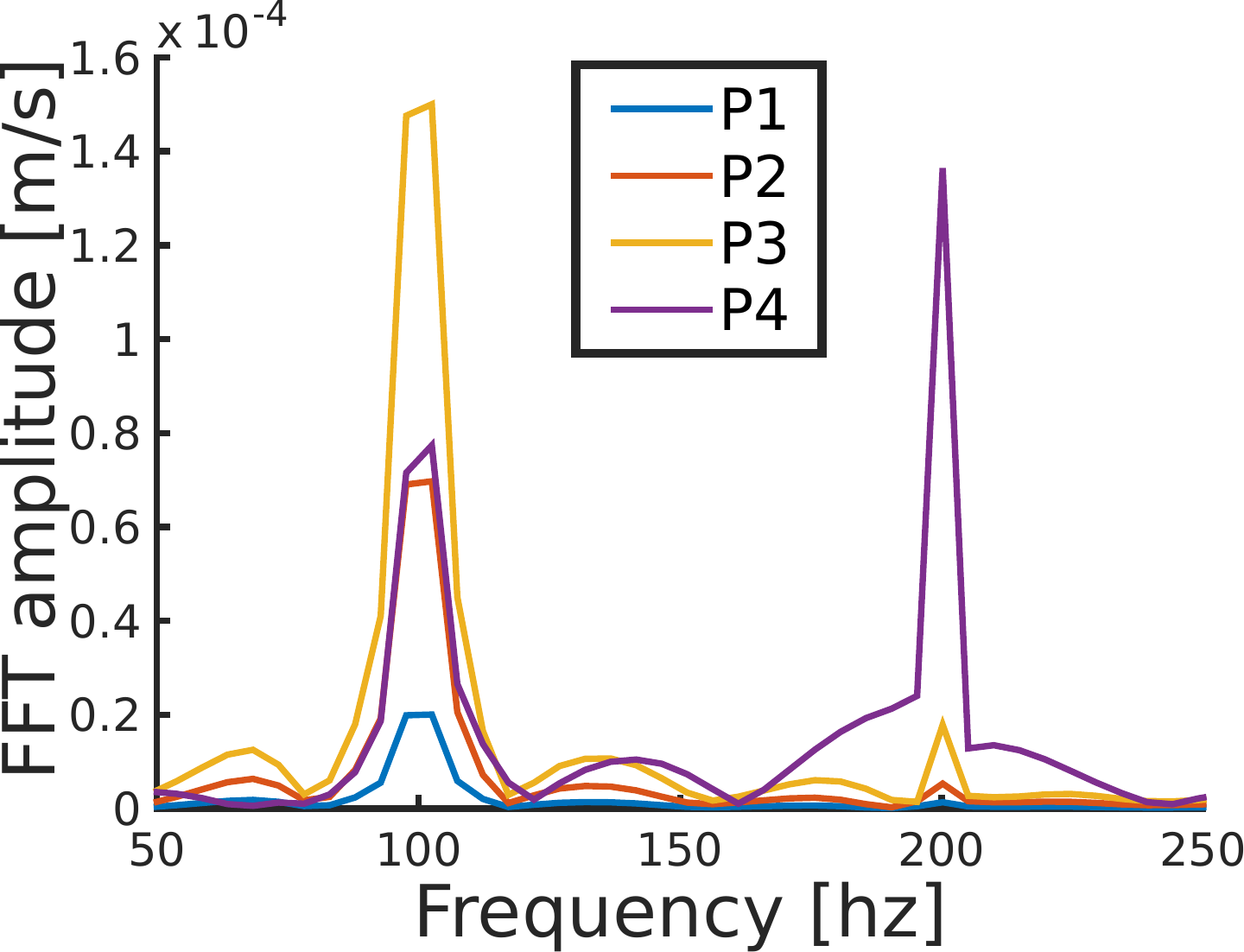}
  \caption{Z axis velocity component.}
  \label{fig:ResSM_Vz}
\end{subfigure}
\hfill
\begin{subfigure}{.42\textwidth}
  \includegraphics[width=\textwidth]{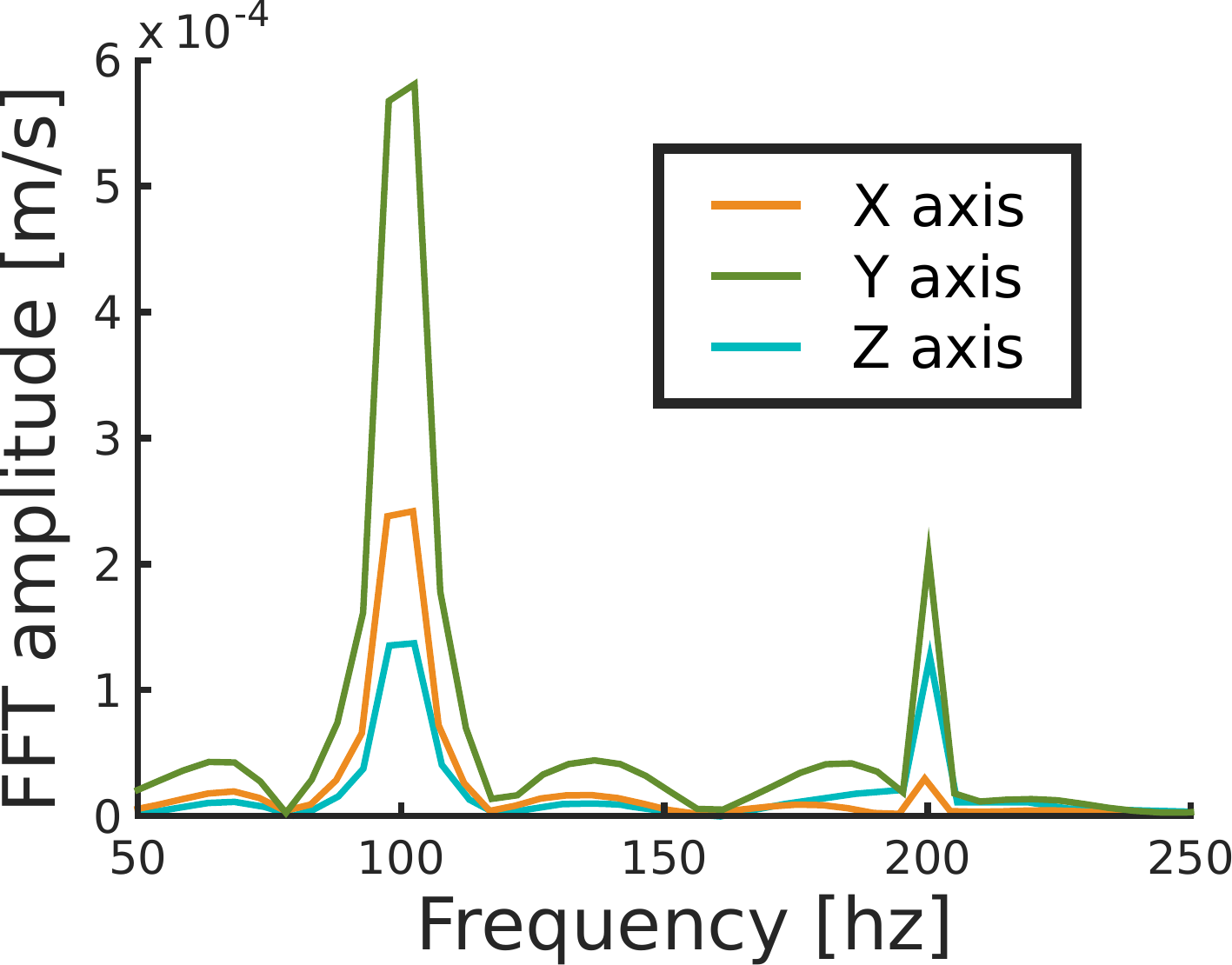}
  \caption{Max. amplitude over each axis.}
  \label{fig:ResSM_MaxAmpl}
\end{subfigure}

  \caption{Single-sided spectrum amplitude of the velocity for probes in the middle of the cochlear duct.  \label{fig:ResSM}}
\end{figure}

These measures (Fig.~\ref{fig:ResSM}) shows that the fluid mainly pulse at the input frequency of 100 hertz with a weak signal at the first harmonic. The strongest fluid flow goes along the Y-axis (Fig.~\ref{fig:ResSM_Vy}) inside the cochlear duct (P4) and at a smaller extend into Corti's organ (P1). This tenuous flux inside Corti's organ is induced by a fluid flow from the main duct through the canal (P2, P3). Indeed, measures along the X and Z axis (Fig.~\ref{fig:ResSM_Vx} and \ref{fig:ResSM_Vz}) show a flow inside the canal having the same magnitude as the one inside Corti's organ. The velocity peak along the Z axis inside the cochlear duct is more difficult to explain. We postulate that this flow could be induced by the reflection of the waves formed by the membrane on the top of the cochlear duct. Moreover, multiple features not taken into account during these simulations, such as the cochlea coiling or the softness of the duct wall and tectorial membrane, would certainly hinder reflection of the fluid flow and thus diminish the harmonics velocity amplitude.

\begin{figure}[ht]
  \center
  \includegraphics[width=\textwidth]{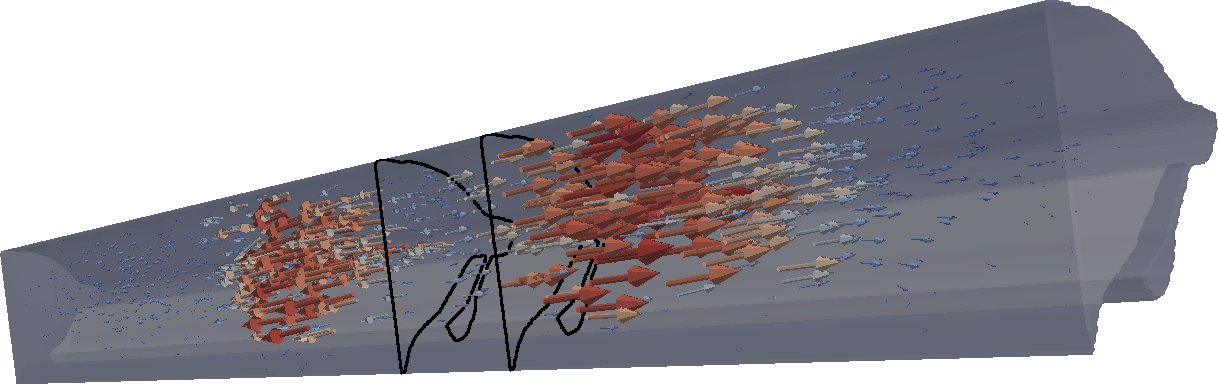}
  \caption{Effect of the membrane deformation, present in between the two black cut, on the fluid velocity. Arrows represents the fluid velocity direction. Their size and color,from small blue to big red, indicate the velocity amplitude.}
  \label{fig:InsideSM}
\end{figure}

These observations were confirmed by volume snapshots of the simulation. Such snapshot contains the pressure and velocity components of each particles in the simulated domain at a given iteration. Using the software ParaView\footnote{ParaView is an open-source, multi-platform data analysis and visualization application. http://www.paraview.org/}, these informations can be visualised in three dimension. Figure~\ref{fig:InsideSM} represents the state of the simulation after 15'000 iterations. The arrows in the duct express the fluid velocity with size and color indicating their amplitude (small and blue means slow, big and red means fast). The two black cross-sectional cuts delimit the current wave evolving along the membrane. This wave appears to push the fluid on both its sides, creating two flows of fluid in opposite direction.

\begin{figure}[ht]
  \center
  \includegraphics[width=\textwidth]{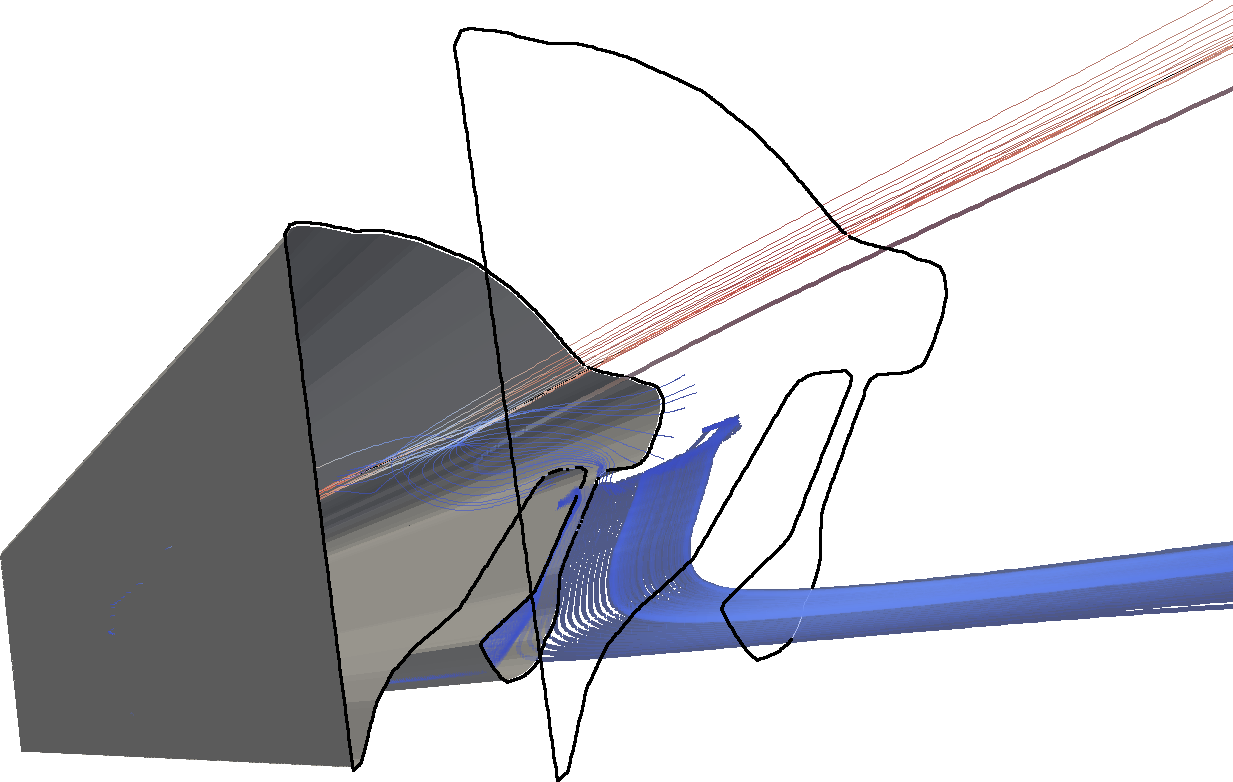}
  \caption{Effect of the membrane deformation, present in between the two black cut, on the fluid path. Streamline represents the path of a particle if it was dropped in the duct. Streamline colors indicate the velocity amplitude of the particles, with blue for slow and red for fast velocity.}
  \label{fig:InsideCorti}
\end{figure}

Figure~\ref{fig:InsideCorti} shows a closer point of view of the current wave emplacement on the membrane. The streamline indicates the path taken by a fluid particle in the duct. Red streamlines indicate higher velocity particles path and are present mainly in the main duct along the Y direction. Blue streamlines show weaker velocity particle paths starting from the wave travelling on the membrane. These streamlines then reaches the opposite side of the main duct (along Z-axis), penetrates through the thin entrance of Corti's organ (along X- and Z-axis) and finally travel along the smaller duct (Y-axis).

These simulations results clearly demonstrated that a fluid flow with low velocity pulsed into Corti's organ in response to an input vibration, more so at the same frequency. As shown in figure~\ref{fig:ScalaMediaSettings}, this input vibration emulated the effect of the ossicles, when excited by sound, on the oval window. Such vibrations then propagated mechanic waves through the perilymph of the vestibular duct inducing an oscillation of the Reissner membrane. This oscillation, preserving the input frequency, created a fluid flow in the cochlear duct that pulsed with low velocity at the exact position of the hair cells.

In this scenario, the Reissner membrane plays a central role by relaying the initial vibrations imposed to the oval window into the cochlear duct. This, by itself, contravene the implicitly and widely accepted scenario in which the Reissner membrane doesn't affect the fluid flow between the vestibular and cochlear duct. Indeed, removing this elastic membrane would strongly impact the dynamic of the fluid flow.

\section{Discussion and Conclusion}

\subsection{Inner ear mechanism}

% Strong opinion 
%a)
One of the inner ear function is to transform the sound that surrounds us into nervous signals. This mechanism is decomposed in multiple steps starting from the vibration of the tympanic membrane. These vibrations are amplified by the ossicles and transmitted to the liquid filled vestibule through the oval windows. The induced fluid waves generate a vibratory response in the cochlea. This vibratory response is attributed to the difference of pressure in the oval and round windows that would create a travelling wave on the basilar membrane. This travelling wave activate then hair cells, or stereocila, present in Corti's organ due to its closeness with the basilar membrane \cite{Yost1989}. 

This simplistic scenario minimizes the potential function of the undervalued Reissner membrane~\cite{Ni2014}. Recently, this membrane has been shown to propagate travelling waves with amplitude comparable to the one of the basilar membrane~\cite{Reichenbach2012}. We simulated this phenomena by applying vibrations at the entrance of the vestibular duct. These vibrations propagated waves inside the duct, thus deforming the Reissner membrane. The oscillation of the membrane maintained the input vibrations frequency and further induced fluid flows in the cochlear duct. Most of the fluid movement went along the membrane, but more importantly, a fluid flow, having half the velocity of the main one, went inside the microscopic organ of Corti. 

While being weaker, this flow pulsed with the same frequency of the input vibrations at the exact location of the sterocilia. These hair cells, depending on their sensibility, could thus have been activated and emitted nervous signal to the brain. Therefore, these results corroborate our initial hypothesis, that movement of the Reissner membrane could be of utmost importance in the mechanism of hearing and yet remains nearly unconsidered nowadays.

In light of the previous results, the ducts structures may play a pivotal role by filtering the perceived sound frequencies. Filtering effects are known to affect waves travelling inside a waveguide of varying cross sectional size~\cite{Kinsler1999}. Cross section measured from the base to the apex of the scala vestibuli and tympani~\cite{Braun2012, Wysocki1999, Yost1989} variates multiple times and form various chambers and bumps. We compared a variety of geometries representing part of these structural features and the ones respecting the proportions of the vestibular duct amplified or attenuated the fluid velocity by more than a twofold factor. These variation in fluid velocity clearly coincided with the structural characteristics of the duct. The variety of structural features of the vestibular and tympanic duct in various mammals~\cite{Wysocki2001} could well impact the range of sound perceived by each species.

\subsection{Simulations limitations and challenges}

While striving to emulate the human ear conditions, our simulations remained far from the reality. While we linked the vestibular and cochlear duct using the Reissner membrane, during the separate simulation steps, the membrane was surrounded on one side by a fluid mimicking the perilymph and by air on the other side. The impact of this inaccurate setting was minimized by increasing the density and friction of the membrane. However, these required simplifications created a gap between our model and the real environment of the cochlea.

Accurately simulating the cochlea poses serious challenges for multiple reasons. One of them resides in the complex structure and anatomy of the inner ear. Despite the recent progress in high-resolution medical imaging, soft tissues such as the Reissner membrane remains hard to observe and measure accurately~\cite{Braun2012, Shibata2009}. Assuming that these issues will be fixed in the future, fully accurate geometries will then cause considerable computational challenges for CFD solvers.

The frequencies range and the scale of geometry details featured in the cochlea leads to hardly tractable computational task. Simulating low and high frequency at the same time requires to chose a time resolution that enable the sampling of the fastest frequency and a run time long enough for the lowest frequency to propagate. Sampling 20 points per period of a 20'000 hertz signal requires a time step of $5 \cdot 10^{-6}$ seconds. With such time step, $10^4$ iterations would be necessary to just generate one period of a 20 hertz signal. The same observation applies to the spatial resolution of the geometry. As example, Corti's organ has geometric features of less than 10 microns, while the unrolled length of the cochlea is greater than 30 millimetres.

The spatial and temporal resolutions of our simulation of one third of the cochlear duct already required a one day runtime with 96 processors to generate enough data. This small simulation, in regard to the entire cochlea, was not even computing forces on the membrane since we applied its previously memorized deformation. Yet tenth of thousands of particles were used to represent accurately the membrane. Adding the resolution of forces on the entire Reissner and basilar membrane would consequently increase the required computational power.

Hopefully, advances in CFD will help address these issues. Multi-domains methods~\cite{Lagrava2012, Touil2014} are a first example of such methods. They aim to facilitate the simulation of geometries showing structure with multiple scales of dimensions by dividing the simulation domain in blocks having each an appropriate spatial resolution. Another issue comes from immersed elastic membrane. Such complex fluid-solid interactions are complex to address with the Lattice-Boltzmann method and are subject to recent advances~\cite{Favier2014}. 

In addition to these CFD improvements, a direct solution to this challenging computational task would be to use the already available computing resources. Palabos is known to manage simulation on clusters having several thousands of processors\footnote{LEMANICUS Blue Gene/Q Supercomputer : http://bluegene.epfl.ch/}.

\subsection{Conclusion}

In conclusion, by simulating the full path of mechanical waves created by vibrations of the oval window, we showed that the Reissner membrane could well play an important role in the activation of hair cells present in the microscopic organ of Corti. In this context, the structure of cochlea ducts presented significant frequency filtering properties by modifying wave velocity by up to a twofold factor.

In order to simulate these phenomenon, we used a highly parallel CFD solver, Palabos. While these experiments where not fully realistic in regard of the human cochlea, they present an important step toward the simulation of inner ear micro- and macro-mechanics. Novel CFD methods coupled with the increasingly accurate measures of the human ear will allow in a near future to conceive truly realistic cochlea models. Given that the required computing resources are already us available for such computational challenge, we only are a few steps away of realising the crucial tool that would help us better understand the human hearing mechanism.

%s\bibliographystyle{plainnat}


\begin{thebibliography}{10}

\bibitem{Bekesy1947}
Békésy, Georg V. “The Variation of Phase Along the Basilar Membrane with Sinusoidal Vibrations.” The Journal of the Acoustical Society of America 19, no. 3 (May 1, 1947): 452–60. doi:10.1121/1.1916502.

\bibitem{Bekesy1960}
Békésy, Georg von, and Ernest Glen Wever. Experiments in Hearing. New York: McGraw-Hill, 1960.

\bibitem{Bohnke1998}
Böhnke, F., and W. Arnold. “Nonlinear Mechanics of the Organ of Corti Caused by Deiters Cells.” IEEE Transactions on Bio-Medical Engineering 45, no. 10 (October 1998): 1227–33.

\bibitem{Braun2012}
Braun, K., F. Böhnke, and T. Stark. “Three-Dimensional Representation of the Human Cochlea Using Micro-Computed Tomography Data: Presenting an Anatomical Model for Further Numerical Calculations.” Acta Otolaryngol 132, no. 6 (2012): 603–13.

\bibitem{Elliot2012}
Elliott, Stephen J., and Christopher A. Shera. “The Cochlea as a Smart Structure.” Smart Materials \& Structures 21, no. 6 (June 2012): 64001. 

\bibitem{Favier2014}
Favier, Julien, Alistair Revell, and Alfredo Pinelli. “A Lattice Boltzmann–Immersed Boundary Method to Simulate the Fluid Interaction with Moving and Slender Flexible Objects.” Journal of Computational Physics 261 (March 15, 2014): 145–61. 

\bibitem{Geisler1995}
Geisler, C. Daniel, and Chunning Sang. “A Cochlear Model Using Feed-Forward Outer-Hair-Cell Forces.” Hearing Research 86, no. 1–2 (June 1995): 132–46.

\bibitem{Givelberg2003}
Givelberg, Edward, and Julian Bunn. “A Comprehensive Three-Dimensional Model of the Cochlea.” Journal of Computational Physics 191, no. 2 (2003): 377–91.

\bibitem{Gray1918}
Gray, Henry. Anatomy of the Human Body. Philadelphia: Lea \& Febiger, 1918. 

\bibitem{Huber2014}
Huber, Christian, Babak Shafei, and Andrea Parmigiani. “A New Pore-Scale Model for Linear and Non-Linear Heterogeneous Dissolution and Precipitation.” Geochimica et Cosmochimica Acta 124 (January 1, 2014): 109–30.

\bibitem{Kemp2002}
Kemp, David T. “Otoacoustic Emissions, Their Origin in Cochlear Function, and Use.” British Medical Bulletin 63, no. 1 (October 1, 2002): 223–41. 

\bibitem{Kinsler1999}
Kinsler, Lawrence E, Austin R Frey, Alan B Coppens, and James V Sanders. “Fundamentals of Acoustics.” Fundamentals of Acoustics, 4th Edition, by Lawrence E. Kinsler, Austin R. Frey, Alan B. Coppens, James V. Sanders, Pp. 560. ISBN 0-471-84789-5. Wiley-VCH, December 1999. 1 (1999).

\bibitem{Lagrava2012}
Lagrava, D., O. Malaspinas, J. Latt, and B. Chopard. “Advances in Multi-Domain Lattice Boltzmann Grid Refinement.” Journal of Computational Physics 231, no. 14 (May 20, 2012): 4808–22. 

\bibitem{Manoussaki2000}
Manoussaki, D., and R. Chadwick. “Effects of Geometry on Fluid Loading in a Coiled Cochlea.” SIAM Journal on Applied Mathematics 61, no. 2 (January 1, 2000): 369–86.


\bibitem{Ni2014}
Ni, Guangjian, Stephen J. Elliott, Mohammad Ayat, and Paul D. Teal. “Modelling Cochlear Mechanics.” BioMed Research International 2014 (July 23, 2014): e150637.

\bibitem{Reichenbach2012}
Reichenbach, Tobias, Aleksandra Stefanovic, Fumiaki Nin, and A. J. Hudspeth. “Waves on Reissner’s Membrane: A Mechanism for the Propagation of Otoacoustic Emissions from the Cochlea.” Cell Rep 1, no. 4 (April 2012): 374–84.

\bibitem{Robles2001}
Robles, Luis, and Mario A. Ruggero. “Mechanics of the Mammalian Cochlea.” Physiological Reviews 81, no. 3 (July 2001): 1305–52.

\bibitem{Shibata2009}
Shibata, Takashi, Sumiko Matsumoto, Tetsuzo Agishi, and Teiko Nagano. “Visualization of Reissner Membrane and the Spiral Ganglion in Human Fetal Cochlea by Micro-Computed Tomography.” American Journal of Otolaryngology 30, no. 2 (2009): 112–20.

\bibitem{Steele2009}
Steele, Charles R., Jacques Boutet de Monvel, and Sunil Puria. “A MULTISCALE MODEL OF THE ORGAN OF CORTI.” Journal of Mechanics of Materials and Structures 4, no. 4 (2009): 755–78.

\bibitem{Touil2014}
Touil, Hatem, Denis Ricot, and Emmanuel Lévêque. “Direct and Large-Eddy Simulation of Turbulent Flows on Composite Multi-Resolution Grids by the Lattice Boltzmann Method.” Journal of Computational Physics 256 (January 1, 2014): 220–33. 

\bibitem{Ulfendahl1997}
Ulfendahl, Mats. “Mechanical Responses of the Mammalian Cochlea.” Progress in Neurobiology 53, no. 3 (October 1997): 331–80. 

\bibitem{Wysocki1999}
Wysocki, Jaros aw. “Dimensions of the Human Vestibular and Tympanic Scalae.” Hearing Research 135, no. 1–2 (1999): 39–46.

\bibitem{Wysocki2001}
Wysocki, Jaros aw. “Dimensions of the Vestibular and Tympanic Scalae of the Cochlea in Selected Mammals.” Hearing Research 161, no. 1 (2001): 1–9.

\bibitem{Yost1989}
Yost, William A, and DW Nielsen. "Fundamentals of Hearing". Holt, Rinehart, and, 1989. 

\bibitem{Zeng2013}
Zeng, Fan-Gang, and Richard R. Fay. Cochlear Implants: Auditory Prostheses and Electric Hearing. Springer Science \& Business Media, 2013.

\bibitem{Zimny2013}
Zimny, Simon, Bastien Chopard, Orestis Malaspinas, Eric Lorenz, Kartik Jain, Sabine Roller, and Jörg Bernsdorf. “A Multiscale Approach for the Coupled Simulation of Blood Flow and Thrombus Formation in Intracranial Aneurysms.” Procedia Computer Science, 2013 International Conference on Computational Science, 18 (2013): 1006–15.


\bibitem{Zwislocki1974}
Zwislocki, J. J. “Cochlear Waves: Interaction between Theory and Experiments.” The Journal of the Acoustical Society of America 55, no. 3 (March 1, 1974): 578–83.
 
\end{thebibliography}
\end{document}